\documentclass[reprint,prb,aps,twocolumn,superscriptaddress,10pt]{revtex4-2}
\usepackage{amsmath}
\usepackage[english]{babel}
\usepackage{graphicx}
\usepackage{color}
\usepackage{sidecap}
\usepackage{siunitx}
\usepackage[dvipsnames]{xcolor}

\usepackage[colorlinks,
            citecolor = blue, 
						linkcolor = blue,
						urlcolor  = blue,
						anchorcolor = blue,
						breaklinks = true
						]{hyperref}

\usepackage{cleveref}

\usepackage{soul}
\usepackage{ amssymb }

\newif\ifdraft
\drafttrue

\def \NIMSKW{Research Center for Electronic and Optical Materials, National Institute for Materials Science, 1-1 Namiki, Tsukuba 305-0044, Japan}
\def \TUM{Technical University of Munich, TUM School of Natural Sciences, Physics Department, 85748 Garching, Germany}
\def \MCQST{Munich Center for Quantum Science and Technology (MCQST), Schellingstr. 4, 80799 M{\"u}nchen, Germany}
\def \NIMSTT{Research Center for Materials Nanoarchitectonics, National Institute for Materials Science,  1-1 Namiki, Tsukuba 305-0044, Japan,\\$^\ddag$These authors contributed equally to this work}

\def \Basel{Department of Physics, University of Basel, Klingelbergstrasse 82, Basel CH-4056, Switzerland}

\begin{document}


\title{Spectroscopy of Wigner crystal polarons in an atomically thin semiconductor}

\author{L.~Wang$^\ddag$}
\affiliation{\Basel}

\author{F.~Menzel$^\ddag$}
\affiliation{\Basel}

\author{F.~Pichler$^\ddag$}
\affiliation{\TUM}
\affiliation{\MCQST}

\author{P.~Kn\"uppel}
\affiliation{\Basel}

\author{~\\ K.~Watanabe}
\affiliation{\NIMSKW}

\author{T.~Taniguchi}
\affiliation{\NIMSTT}

\author{M.~Knap}
\affiliation{\TUM}
\affiliation{\MCQST}

\author{T.~Smole\'nski}
\email{tomasz.smolenski@unibas.ch}
\affiliation{\Basel}

\maketitle

{\bf Strongly interacting electrons in two-dimensional systems can spontaneously break translational symmetry, forming a periodic Wigner crystal~\cite{wigner1934}. Although these crystals have been realized in several platforms~\cite{lozovik1975,grimes1979,Andrei1988,Goldman1990,Williams1991,yoon1999,Ye2002,spivak2004,spivak2010,deng2016,shapir2019,falson2022,Pack2024,xiang2025,zhang2025,smolenski2021,zhou2021,sung2025,Chen2025}, experimental studies of their collective many-body excitations in the absence of a magnetic field~\cite{Chen2025} remain an outstanding challenge. Here, we access this regime optically by uncovering Wigner crystal polarons: novel light-matter excitations arising from the dressing of excitons by collective excitations of the Wigner crystal. These hybrid quasiparticles manifest as new optical resonances in cryogenic reflectance spectra of a charge-tunable WSe$_2$ monolayer, appearing concurrently with previously identified exciton umklapp transitions~\cite{smolenski2021,shimazaki2021}. In contrast to the latter, the energies of Wigner crystal polarons are governed not only by the electronic lattice constant but also by their hybridization with attractive exciton-polarons, whose strength is controlled by electronic interactions. These novel many-body excitations provide an optical interface to the spin state of the Wigner crystal, which as we demonstrate, can be controlled both magnetically and optically. Our work establishes layered materials as a unique platform for exploring dynamical impurity dressing by strongly correlated electronic orders.}

Understanding the ground states and collective excitations of strongly correlated electronic phases lies at the heart of condensed matter physics. A paradigmatic example of such a phase is the Wigner crystal (WC), which forms when Coulomb interactions dominate over kinetic energy in a two-dimensional (2D) electron system~\cite{wigner1934}. Following studies in conventional semiconductors, where WCs are typically characterized via transport properties~\cite{lozovik1975,grimes1979,Andrei1988,Goldman1990,Williams1991,yoon1999,Ye2002,spivak2004,spivak2010,deng2016,shapir2019,falson2022}, interest in WCs has been recently reignited by the advent of van der Waals (vdW) heterostructures~\cite{geim2013,manzeli2017} based on transition metal dichalcogenide (TMD) monolayers. With their gate-tunability, weak dielectric screening, and flat electronic bands, these structures not only provide a tunable platform for realizing 
WCs~\cite{smolenski2021,zhou2021,sung2025,Chen2025,Pack2024,xiang2025,zhang2025}, but also uniquely enable probing them with excitons---fundamental optical excitations that act as robust bosonic impurities in a fermionic environment~\cite{Sidler2017}. In particular, exciton scattering off the periodic repulsive potential generated by the WC gives rise to an umklapp resonance whose energy serves as a quantitative probe of the WC lattice constant~\cite{smolenski2021,shimazaki2021}. Yet, the opposite regime---where the excitons strongly couple to collective excitations of the WC---has thus far remained elusive. This is the realm of polarons: quasiparticles arising from dressing an impurity with excitations of a surrounding medium~\cite{Schirotzek2009,Kohstall2012,koschorreck2012,Sidler2017}, originally introduced to describe electrons interacting with phonons in real crystals~\cite{Landau1933}. 
Although recent experiments have begun to explore how correlated phases affect exciton–polarons, in particular in the presence of moir\'e potentials~\cite{liu2021a,kiper2025,evrard2025,upadhyay2024,yan2024}, the nature of polaron dressing by collective excitations in spontaneously ordered electronic crystals---and the possible emergence of novel quasiparticles---remains unresolved.

Here, we report the direct observation of \textit{Wigner crystal polarons} (WPs): novel light-matter excitations in the resonant optical response of a WSe$_2$ monolayer hosting a robust WC at low temperature $T$ and electron density $n_\mathrm{e}$. We show that WPs manifest as umklapp replicas of two attractive exciton-polarons (APs) and appear within the same $T-n_\mathrm{e}$ phase space as the exciton umklapp resonance. In stark contrast to the latter, both WPs exhibit a sizable blueshift relative to the corresponding APs even in low-density limit, indicating that they arise not only from Bragg scattering of mobile excitons, but also from their dressing by collective excitations of the WC. We show that the intensities of WPs are determined by the spin state of the WC, which we control not only with an external magnetic field but also optically, using circularly polarized light. Our results demonstrate the potential of vdW materials for characterizing emergent quasiparticle excitations arising from strong interactions between excitons and charge-ordered electronic phases.

\begin{figure*}[]
	\includegraphics[width=\textwidth]{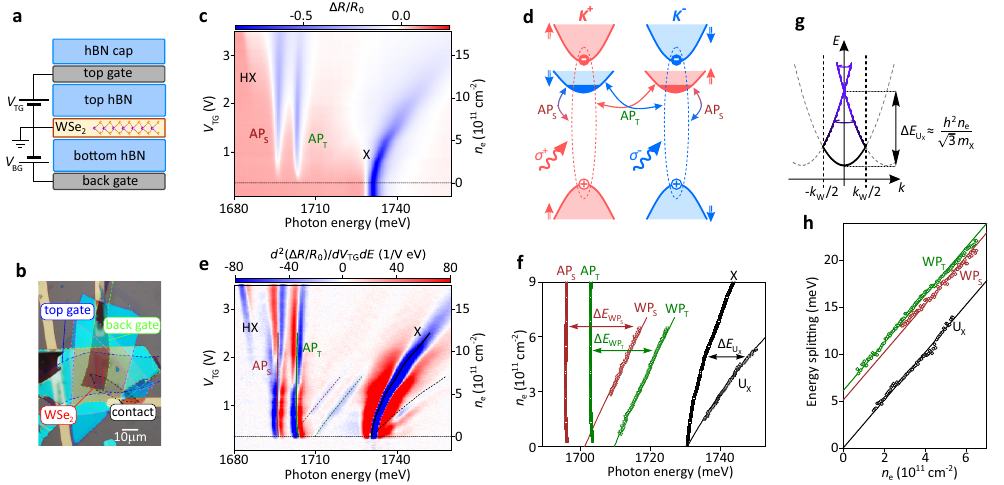}
	\caption{{\bf Optical signatures of a Wigner crystal in a WSe$_2$ monolayer.} ({\bf a,b}) Schematic~({\bf a}) and optical micrograph~({\bf b}) of the charge-tunable WSe$_2$ monolayer device explored in the main text (see Methods Sec.~1 for the fabrication details). ({\bf c})~Reflectance contrast spectrum as a function of the electron density $n_\mathrm{e}$ at $B=0$ and $T=1.6$~K. ({\bf d})~Schematic illustrating the band structure of a WSe$_2$ monolayer as well as valley-selective optical selection rules. ({\bf e}) $n_\mathrm{e}$-evolution of the reflectance contrast differentiated with respect to the gate voltage and photon energy. Solid lines represent the fitted energies of exciton and both AP transitions, while dashed lines mark the corresponding umklapp and WP resonances originating from scattering off the Wigner crystal. ({\bf f}) Energies of X, AP$_\mathrm{T}$, and AP$_\mathrm{S}$ transitions as well as the corresponding umklapp and WP peaks extracted from the data in ({\bf c,e}) in the low electron density limit (see Methods Sec.~4 for details). ({\bf g}) Illustration of the back-folded exciton band structure, giving rise to a blueshifted-umklapp resonance (purple). ({\bf h}) $n_\mathrm{e}$-dependent energy splittings between each main peak and the corresponding umklapp transition. Each dataset is fitted with a linear dependence $\Delta E=h^2n_\mathrm{e}/\sqrt{3}m_\mathrm{X}+\Delta$ with fixed $m_\mathrm{X}\approx0.68m_0$ and different offset $\Delta$: 0 for U$_\mathrm{X}$, 6~meV for WP$_\mathrm{T}$, and 5~meV for WP$_\mathrm{S}$.\label{fig:Fig1}}
\end{figure*}

Our experiments are performed on two dual-gated WSe$_2$ monolayer devices, yielding consistent results. Their optical response is explored in a confocal microscope setup integrated with a magneto-optical cryostat enabling reaching sample temperatures down to $T=1.6$~K (Methods Sec.~2). In the main text, we focus on the results obtained from device A, whose schematic and optical micrograph are depicted in Figs.~\ref{fig:Fig1}{\bf a,b} (see Methods Sec.~10 for device~B). 

Fig.~\ref{fig:Fig1}{\bf c} shows low-temperature resonant reflectance contrast spectrum ($\Delta R/R_0$) of our device as a function of electron density $n_\mathrm{e}$, whose value is controlled by applying a top gate voltage $V_\mathrm{TG}$. The dependence of $n_\mathrm{e}$ on $V_\mathrm{TG}$ is precisely calibrated based on the observation of Landau level (LL) signatures at high magnetic fields~\cite{Wang2017,Smolenski2019,Li2020} (Methods Sec.~6). Consistent with previous reports~\cite{Ross2013,jones2013}, we find that the oscillator strength of the exciton (X) resonance is progressively transferred upon electron doping to two distinct red-shifted AP resonances: AP$_\mathrm{S}$ and AP$_\mathrm{T}$. They arise from attractive interactions between optically-created excitons and the electrons residing in, respectively, the same and opposite valleys (see Fig.~\ref{fig:Fig1}{\bf d}), which are responsible for the formation of singlet (intravalley) and triplet (intervalley) trions~\cite{courtade2017}. Even though the X resonance in this regime transforms into a repulsive polaron, for simplicity we refer to it as X, since we focus on the low-density regime. At higher densities $\gtrsim10^{12}\ \mathrm{cm}^{-2}$, we observe a third resonance at even lower energies, which is due to the formation of hexcitons (HX)~\cite{VanTuan2022} originating from simultaneous exciton dressing by electrons in both valleys.

\begin{figure*}[]
	\includegraphics[width=\textwidth]{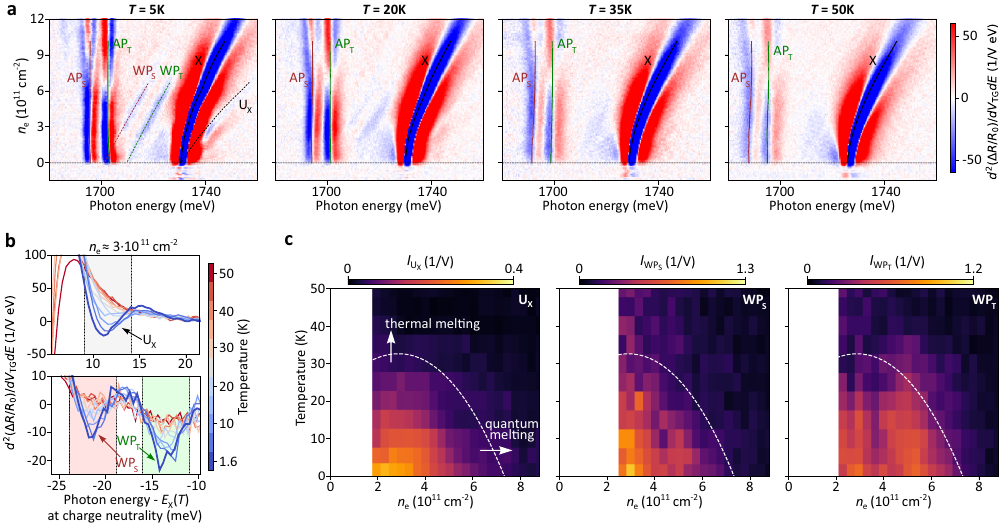}
	\caption{{\bf Robustness of the Wigner crystal to thermal and quantum fluctuations.} ({\bf a}) Electron-density evolutions of differentiated reflectance contrast spectra acquired at various temperatures (as indicated). Solid lines mark the fitted energies of main X, AP$_\mathrm{T}$, and AP$_\mathrm{S}$ resonances, while dashed lines represent expected positions of the umklapp peaks at $T=5$~K. ({\bf b}) Linecuts through the plots in ({\bf a}) (as well as similar datasets from different temperatures) showing temperature-dependent spectral profiles of the exciton umklapp (top) and Wigner crystal polarons (bottom) at $n_\mathrm{e}\approx3\cdot10^{11}\ \mathrm{cm}^{-2}$. To account for the temperature-induced bandgap shift, the energy axes of all plots are offset by the exciton energy $E_\mathrm{X}(T)$ at charge neutrality. The shaded areas mark a-few-meV-wide energy ranges around the respective umklapp resonances used for determining their spectral weight by averaging background-corrected $|d^2(\Delta R/R_0)/dV_\mathrm{TG}dE|$ signal (see Methods Sec.~8 for details). ({\bf c}) $T$--$n_\mathrm{e}$ phase diagrams of the Wigner crystal acquired using U$_\mathrm{X}$ (left), WP$_\mathrm{S}$ (middle), and WP$_\mathrm{T}$ (right). Each map shows averaged spectral weight of the umklapp resonance divided by the oscillator strength of the corresponding main optical transition. The domed-shaped phase boundary (identical for all plots; dashed line) is a guide to the eye.\label{fig:Fig2_phase_diag}}
\end{figure*}

Along with these well-established optical resonances, our $\Delta R/R_0$ spectra also exhibit a set of weaker features in the low-$n_\mathrm{e}$ regime, where electrons are expected to form a WC~\cite{Pack2024,xiang2025,smolenski2021,zhou2021,sung2025,Chen2025,zhang2025}. These transitions become readily discernible upon differentiating the data with respect to both $V_\mathrm{TG}$ and photon energy ($E$), as shown in Fig.~\ref{fig:Fig1}{\bf e} (see Methods Sec.~3). In particular, at high energies we observe the U$_\mathrm{X}$ peak that blueshifts relative to the X, with the splitting $\Delta E_\mathrm{U_X}$ scaling proportionally to $n_\mathrm{e}$ (Figs.~\ref{fig:Fig1}{\bf f,h}). This is an umklapp signature of the WC~\cite{smolenski2021} arising due to Bragg scattering of the excitons by the periodic WC potential (Fig.~\ref{fig:Fig1}{\bf g}), which leads to a new optical excitation at $\Delta E_\mathrm{U_X}\approx\hbar^2k_\mathrm{W}^2/2m_\mathrm{X}=h^2n_\mathrm{e}/\sqrt{3}m_\mathrm{X}$, where $k_\mathrm{W}$ is the reciprocal lattice constant of the triangular WC and $m_\mathrm{X}$ is the exciton mass. Thus far, such an umklapp resonance has been observed exclusively in MoSe$_2$-based systems~\cite{smolenski2021,zhou2021,sung2025,Chen2025,shimazaki2021,kiper2025}. Compared to MoSe$_2$ monolayers~\cite{smolenski2021}, U$_\mathrm{X}$ in the present case persists up to about twice higher densities, $n_\mathrm{c}\sim7\cdot10^{11}\ \mathrm{cm}^{-2}$, and exhibits almost twice larger $\Delta E_\mathrm{U_X}$ splitting at a given $n_\mathrm{e}$, indicating lower $m_\mathrm{X}$. Both effects can occur simultaneously because, unlike MoSe$_2$, in WSe$_2$ the electrons forming optically active excitons and those forming the WC reside in different spin-orbit-split bands (cf.~Fig.~\ref{fig:Fig1}{\bf d}). The excitonic transitions involve a more dispersive top conduction band, with a theoretically-predicted effective mass of $m_{\mathrm{t},\mathrm{e}}^*\approx0.3m_0$~\cite{Kormanyos2015}, where $m_0$ is the free electron mass. Together with the hole mass $m_\mathrm{h}^*\approx0.4m_0$~\cite{Kormanyos2015,Fallahazad2016}, this yields relatively low translational exciton mass $m_{\mathrm{t},\mathrm{e}}^*+m_\mathrm{h}^*\approx0.7m_0$, which remains in excellent agreement with $m_\mathrm{X}\approx0.68m_0$ extracted from our data in Fig.~\ref{fig:Fig1}{\bf h}, that is further confirmed by prior electron scattering experiments~\cite{Hong2020}. By contrast, the WC electrons originate from less-dispersive bottom conduction band with a larger effective mass $m_{\mathrm{b},\mathrm{e}}^*>m_{\mathrm{t},\mathrm{e}}^*$, which supports the WC formation. Assuming that its melting---corresponding to the disappearance of U$_\mathrm{X}$---occurs when the ratio of Coulomb to kinetic energies $r_\mathrm{s}\sim m_{\mathrm{b},\mathrm{e}}^*/\sqrt{n_\mathrm{e}}\approx30$, we obtain $m_{\mathrm{b},e}^*\approx m_0$ that is larger than theoretical predictions~\cite{Kormanyos2015}, suggesting a role of disorder in stabilizing the WC at a reduced~$r_s$~\cite{xiang2025}. 

Notably, differentiated spectra in Fig.~\ref{fig:Fig1}{\bf e} also display two other weak blueshifting resonances in the energy range between AP and X, which remained elusive in previous studies of the MoSe$_2$ platform. The positions of these resonances, WP$_\mathrm{S}$ and WP$_\mathrm{T}$, are traced in Fig.~\ref{fig:Fig1}{\bf f}. The energy splitting between them is essentially independent of the electron density, and remains close to the exchange-induced splitting $\sim7$~meV between AP$_\mathrm{S}$ and AP$_\mathrm{T}$ resonances. Moreover, both peaks disappear around the same critical density $n_\mathrm{c}\sim7\cdot10^{11}\ \mathrm{cm}^{-2}$ as the exciton umklapp transition, suggesting that they likewise originate from umklapp scattering of the WC. We first rule out singlet and triplet trions undergoing such an umklapp process. In this scenario, the energy splitting $\Delta E_\mathrm{WP_\mathrm{S,T}}$ between a given WP$_\mathrm{S,T}$ umklapp branch and the corresponding main AP$_\mathrm{S,T}$ resonance should increase with $n_\mathrm{e}$ with a slope inversely proportional to the translational trion mass $m_\mathrm{T}=m_{\mathrm{b},\mathrm{e}}^*+m_\mathrm{X}$. Owing to the aforementioned difference in effective masses between the two spin-orbit-split conduction bands, such an $m_\mathrm{T}$ is expected to be significantly larger than $m_\mathrm{X}$, suggesting considerably lower slope of the associated trion umklapp splitting. By contrast, $\Delta E_\mathrm{WP_\mathrm{S,T}}$ in our experiments exhibits a slope that is similar to $\Delta E_\mathrm{U_\mathrm{X}}$, and can be reproduced assuming the same translational mass of $m_\mathrm{X}\approx0.68m_0$, as shown by the fits in Figs.~\ref{fig:Fig1}{\bf f,h}. This observation suggests that WP$_\mathrm{S}$ and WP$_\mathrm{T}$ transitions rather arise due to umklapp-scattered neutral excitons that are dressed into \emph{Wigner crystal polarons} by the WC.

\begin{figure*}[]
    \includegraphics[width=\textwidth]{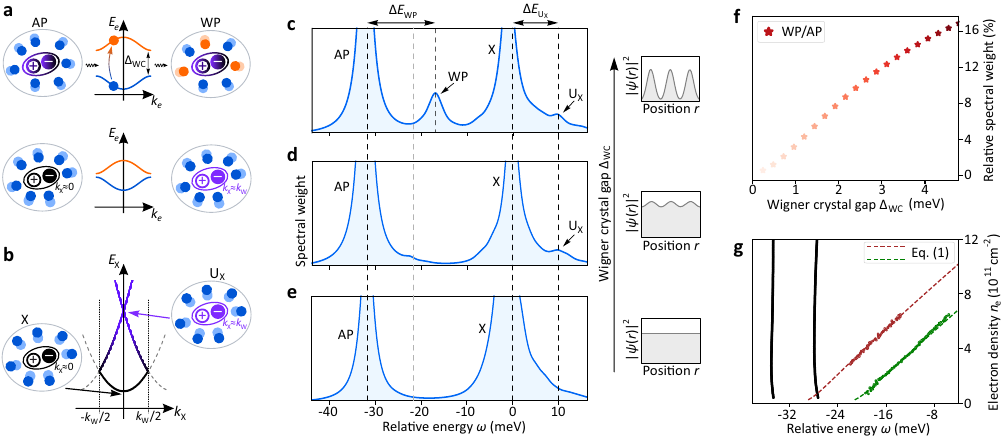}
    \caption{{\bf Theoretical description of the Wigner crystal polaron.} ({\bf a}) Illustration of an attractive and Wigner crystal polaron formation. In the weak WC limit with a small gap $\Delta_\mathrm{WC}$ (bottom), the AP (WP) originates primarily from particle-hole dressing of a ground (Bragg-scattered) exciton with momentum $k\approx0$ ($k\approx k_\mathrm{W}$). For larger $\Delta_\mathrm{WC}$ (top), these two excitations hybridize due to strong attractive exciton-electron interactions, mediated by particle-hole excitations across the WC gap. The hybridization results in an enhancement of the WP spectral weight and an increase of its energy detuning from the AP. ({\bf b}) Cartoon showing the corresponding exciton umklapp scattering by repulsive exciton-electron interactions, where the energy splitting between the the main and umklapp branches is largely independent of $\Delta_\mathrm{WC}$. (\textbf{c}-\textbf{e}) Exciton spectral functions calculated using a spin-independent model within the Chevy approximation for a fixed $n_\mathrm{e}=5\cdot10^{11}\ \mathrm{cm}^{-2}$, and various WC gaps: ({\bf e}) $\Delta_\mathrm{WC}=0$; ({\bf d}) weak WC with a small gap where U$_\mathrm{X}$ appears at an energy detuning $\Delta E_{\mathrm{U}_\mathrm{X}} \approx \hbar^2k_\mathrm{W}^2/2m_\mathrm{X}$; ({\bf c}) strong WC with a large gap, where WP of sizable spectral weight appears at an energy splitting $\Delta E_{\mathrm{WP}} > \Delta E_{\mathrm{U}_\mathrm{X}}$. (\textbf{f}) The relative spectral weight of the WP compared to the AP calculated as a function of $\Delta_{\mathrm{WC}}$ for a fixed $n_\mathrm{e}=5\cdot10^{11}$ cm$^{-2}$. (\textbf{g}) Electron-density dependent energies of both singlet and triplet AP and WP transitions: points represent the experimental data from Fig.~\ref{fig:Fig1}, while dashed lines mark the fitted energies of WP$_\mathrm{S}$ and WP$_\mathrm{T}$ resonances using a single-parameter effective model given by Eq.~\eqref{eq:hybridization}. For each WP branch, the energy is obtained by adding the experimentally-determined AP energy to the theoretically calculated AP-WP energy splitting at a given $n_\mathrm{e}$.}
	\label{fig:Fig_theory}
\end{figure*}

In contrast to the previously observed exciton umklapp peak U$_\mathrm{X}$, the energy splittings $\Delta E_\mathrm{WP_\mathrm{S,T}}$ of both WPs do not extrapolate to zero in the low-density limit, but instead exhibit a sizable offset $\Delta$ (see Fig.~\ref{fig:Fig1}{\bf h}). Before addressing the origin of this surprising effect and discussing the nature of the WP$_\mathrm{S,T}$ resonances, we first provide further evidence for their identification. To this end, we investigate melting of the WC by tracking $n_\mathrm{e}$-evolutions of differentiated spectra at elevated temperatures, where both U$_\mathrm{X}$ and WP$_\mathrm{S,T}$ transitions weaken considerably (Fig.~\ref{fig:Fig2_phase_diag}{\bf a}). For quantitative comparison, we extract their spectral weights by integrating the background-corrected $|d^2(\Delta R/R_0)/dV_\mathrm{TG}dE|$ signal in narrow energy windows around each peak (Fig.~\ref{fig:Fig2_phase_diag}{\bf b} and Methods Sec.~8). The resulting amplitudes are normalized by the oscillator strenghts of the corresponding X or AP resonances, and plotted as a function of both $T$ and $n_\mathrm{e}$ in Fig.~\ref{fig:Fig2_phase_diag}{\bf c} with identical color scales up to an overall multiplicative factor. Strikingly, each umklapp branch is observed within similar $T-n_\mathrm{e}$ phase space regions, reminiscent of the characteristic dome-shaped phase diagram of the WC~\cite{spivak2004}. Around the critical density $n_c \sim 7\cdot10^{11}\,\mathrm{cm}^{-2}$, the electron kinetic energy $\sim1$~meV greatly exceeds thermal energy $\sim 0.1~\mathrm{meV}$ at $T = 1.6~\mathrm{K}$, leading to quantum melting of the WC for $n_\mathrm{e} \gtrsim n_\mathrm{c}$. At higher $T$, thermal fluctuations of the electrons reduce the $n_\mathrm{e}$ range of the WC phase, which ceases to exist above the critical temperature $T_\mathrm{m}\approx30$~K. This value is about three times larger than that of the WC in a MoSe$_2$ monolayer~\cite{smolenski2021,zhou2021,sung2025,Chen2025}, consistent with the enhanced stability of the WC in the WSe$_2$ system.

The above observations unequivocally confirm the connection between electronic crystallization and the emergence of the WP resonances in the optical spectra. The mechanism behind their formation is, however, qualitatively distinct from simple Bragg reflection of the excitonic bands that leads to the U$_\mathrm{X}$ resonance. Specifically, we find that WPs---similarly as the APs---arise due to strong attractive interactions between the electrons and excitons. As illustrated in Figs.~\ref{fig:Fig_theory}\textbf{a,b}, these interactions mediate a hybridization between two polaron branches originating from dressing of the main and Bragg-scattered excitons. The formation of the WP, therefore, involves both exciton Bragg-umklapp scattering and coupling to particle-hole excitations across the many-body WC gap $\Delta_\mathrm{WC}$ (that should be understood as the mean-field order parameter of the WC). This sets the scale of the effective potential felt by the electrons, and is responsible for the presence of the sizable offset between the AP and WP energies in the low-density limit (Fig.~\ref{fig:Fig1}{\bf f,h}).

\begin{figure*}[]
    \includegraphics[width=\textwidth]{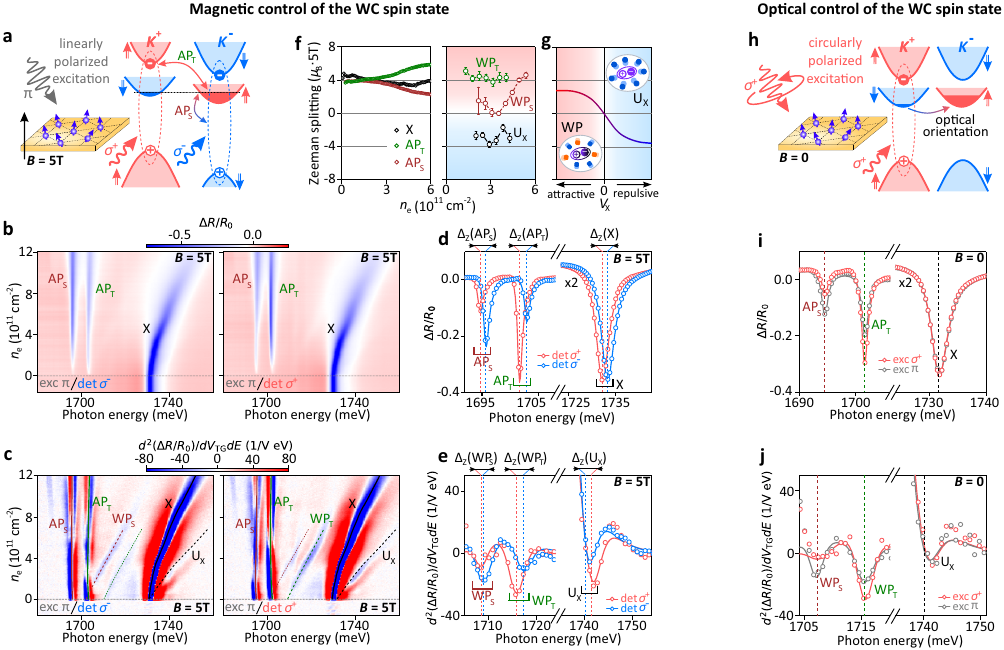}
	\caption{{\bf Magnetic and optical control of the Wigner crystal spin state.}
    ({\bf a,h}) Schematics illustrating two approaches for inducing sizable spin-valley polarization of the Wigner crystal: by applying a magnetic field ({\bf a}) or by illuminating the sample with circularly polarized light at $B=0$ ({\bf h}). ({\bf b,c})~Electron density evolutions of reflectance contrast ({\bf b}) and its second derivative ({\bf c}). The spectra were measured under linearly polarized excitation in two circular polarizations of detection at $B=5$~T and $T=1.6$~K. Solid lines show the fitted energies of the main optical transitions, while dashed lines mark expected positions of the umklapp resonances determined by fitting their splitting from the corresponding co-polarized main exciton transitions with $h^2n_\mathrm{e}/\sqrt{3}m_\mathrm{X}+\Delta$ for a fixed $m_\mathrm{X}=0.68m_0$ and different offsets $\Delta$. ({\bf d,e}) Linecuts through the maps in ({\bf b,c}) at $n_\mathrm{e}\approx3\cdot10^{11}\ \mathrm{cm}^{-2}$ showing polarization-resolved main resonances ({\bf d}) and the corresponding umklapp transitions ({\bf e}). The data in ({\bf e}) were binned along the energy axis. Solid lines show the fitted spectral profiles, while dashed ones mark the extracted peak energies (see Methods Sec.~4 for the details of the fitting procedures). ({\bf f}) Zeeman splittings between $\sigma^-$ and $\sigma^+$ polarized branches of the main excitons (left) and of the corresponding umklapp resonances (right) determined as a function of electron density at $B=5$~T. ({\bf g}) Calculated Zeeman splitting of the umklapp-scattered exciton as a function of exciton-electron interaction strength $V_\mathrm{X}$ (see Methods Sec.~13). The splitting is positive (negative) for attractive (repulsive) interactions, consistent with experimental results. ({\bf i,j}) Similar spectra as in ({\bf d,e}) at the same $n_\mathrm{e}$, but acquired at $B=0$ and two different polarizations of excitation: $\sigma^+$ and linear. The difference in intensities of triplet and singlet branches of AP/WP transitions demonstrate optical spin pumping of the WC (cf. Methods Sec.~9).}
	\label{fig:Fig3_Bfield}
\end{figure*}

To obtain an effective understanding of the resulting optical spectra, we build on Refs.~\cite{amelioPolaronFormationInsulators2024, pichler2025purelyelectronicmodelexcitonpolaron}, and describe the polaron dressing within the Chevy approximation~\cite{chevyUniversalPhaseDiagram2006}. We assume spin-independent density-density interactions between electrons and excitons (see Methods Secs.~11~and~14 for details). The WC is treated on a mean-field level, with $\Delta_\mathrm{WC}$ serving as a tuning parameter. As expected, for $\Delta_\mathrm{WC}=0$ (i.e., in the absence of a WC), our model yields two resonances---the X and AP (Fig.~\ref{fig:Fig_theory}\textbf{e}). For small $\Delta_\mathrm{WC}>0$, an exciton umklapp U$_\mathrm{X}$ appears in the calculated spectra (Fig.~\ref{fig:Fig_theory}\textbf{d}) at an energy detuning $\Delta E_\mathrm{U_X}$ largely determined by the reciprocal WC lattice constant: this is a weak WC regime explored in previous studies of MoSe$_2$ monolayers, where only an U$_\mathrm{X}$, but no WP is observed~\cite{smolenski2021,zhou2021,sung2025,Chen2025}. By contrast, we find that the WP becomes discernible only for larger $\Delta_{\mathrm{WC}}$ (Fig.~\ref{fig:Fig_theory}\textbf{c}). Moreover, while $\Delta E_\mathrm{U_X}$ remains largely insensitive to $\Delta_\mathrm{WC}$ in this regime (for a given $n_\mathrm{e}$), both the AP--WP splitting and the relative spectral weight of the WP increase substantially with $\Delta_\mathrm{WC}$ (Figs.~\ref{fig:Fig_theory}\textbf{c},\textbf{f}). These predictions are consistent with our experimental data, explaining not only the larger energy detuning of the WP resonances but also their sizable oscillator strengths, which we estimate to be $\sim$10–20\% of the APs based on spectra measured at low $T$ and $n_\mathrm{e}$ (see Methods Sec.~5 and Extended Data Fig.~\ref{fig:TM_fits}\textbf{e}).

These theoretical observations can be captured by a simplified analytical model describing interaction-mediated hybridization between AP and WP resonances:
\begin{equation}
    H_\mathrm{hyb} = \begin{pmatrix}
        E_\mathrm{AP} && V_\text{hyb}(n_\mathrm{e}) \\
        V^*_\text{hyb}(n_\mathrm{e}) && E_\mathrm{AP} + \frac{h^2}{\sqrt{3}} \frac{n_\mathrm{e}}{m_\mathrm{X}}
    \end{pmatrix}, \label{eq:hybridization}
\end{equation}
where the diagonal terms represent bare AP and WP energies, and $V_\text{hyb}(n_\mathrm{e})$ denotes the hybridization strength, which scales linearly with exciton-electron interactions and increases for larger $\Delta_\mathrm{WC}$. 
Since the overall AP energy $E_\mathrm{AP}$ is difficult to predict theoretically, we focus on the splitting between AP and WP.
As shown in Fig.~\ref{fig:Fig_theory}\textbf{g}, the minimal model Eq.~\eqref{eq:hybridization} reproduces the measured $n_\mathrm{e}$ dependence of the splittings $\Delta E_{\mathrm{WP}_{\mathrm{S,T}}}$ using a single free parameter (see Methods Sec.~12). 
At low densities, the splitting is primarily determined by the hybridization gap $\sim |V_\mathrm{hyb}(n_\mathrm{e})|$, which grows rapidly with $n_\mathrm{e}$ in the experimentally inaccessible regime $n_\mathrm{e} \lesssim 10^{11}\mathrm{cm}^{-2}$, resulting in an apparent offset $\Delta$ at $n_\mathrm{e} = 0$. At higher densities, the diagonal Bragg-scattering term $\sim n_\mathrm{e}/m_\mathrm{X}$ dominates, leading to a linear increase of the splitting with $n_\mathrm{e}$.

Having elaborated on the nature of WP transitions, we now explore their spectral response when the WC becomes spin-valley polarized. To induce such polarization, we pursue two different approaches. In the first one, we apply an external magnetic field, which lifts the degeneracy of the two conduction bands (Fig.~\ref{fig:Fig3_Bfield}{\bf a}) due to the valley Zeeman effect ~\cite{Srivastava2015,Aivazian2015}. Figs.~\ref{fig:Fig3_Bfield}{\bf b,c} show circular-polarization-resolved evolutions of bare and differentiated reflectance spectra at $B=5$~T. Owing to intra- and intervalley dressing of AP$_\mathrm{S}$ and AP$_\mathrm{T}$ resonances, the large field-induced electronic spin orientation manifests as a sizable circular polarization degree of the APs, with AP$_\mathrm{S}$ (AP$_\mathrm{T}$) stronger in $\sigma^-$ ($\sigma^+$) helicity, as seen clearly in the linecut of Fig.~\ref{fig:Fig3_Bfield}{\bf d}. Strikingly, WP$_\mathrm{S}$ and WP$_\mathrm{T}$ resonances exhibit the same behavior as the respective APs (Fig.~\ref{fig:Fig3_Bfield}{\bf e}), further confirming that they arise from polaron dressing of excitons by electrons in the same and opposite valleys, respectively. 

In the second approach, the spins of WC electrons are optically oriented with circularly polarized light (Fig.~\ref{fig:Fig3_Bfield}{\bf h}), giving rise to a similar change in the WP intensities even at $B=0$ (Figs.~\ref{fig:Fig3_Bfield}{\bf i,j}). As discussed in Methods Sec.~9, this optical spin-pumping method~\cite{robert2021} enables sizable spin-polarization degrees of the WC without affecting its charge-order parameter, thereby paving the way for future dynamical spin-control experiments.

Independently of whether the WC is magnetically or optically oriented, the energy detunings $\Delta E_\mathrm{WP_\mathrm{S,T}}$ of both WP branches from the corresponding AP transitions exhibit very similar doping dependence to that in the spin-unpolarized case. This observation indicates that the singlet-triplet fine structure of exciton-polarons plays no significant role in the WP formation. However, unlike the zero-field case, the optical resonances under $B=5$~T exhibit finite Zeeman splittings. In particular, the splittings of both WPs have the same sign as those of the X, AP$_\mathrm{S}$, and AP$_\mathrm{T}$ resonances (see Figs.~\ref{fig:Fig3_Bfield}{\bf d-f} and Extended Data Fig.~\ref{ex:zeeman}). This contrasts sharply with the U$_\mathrm{X}$, for which the energy order of $\sigma^+$ and $\sigma^-$-polarized branches---and hence the sign of the Zeeman splitting---is reversed (Figs.~\ref{fig:Fig3_Bfield}{\bf e,f}). For the main excitonic transitions, the splitting is governed predominantly by the Zeeman effect of the bright neutral exciton, yielding $\Delta_\mathrm{Z}\approx4\mu_BB$ (where $\mu_B$ denotes the Bohr magneton)~\cite{Srivastava2015,Aivazian2015}. Conversely, the umklapp transitions are expected to exhibit smaller $\Delta_\mathrm{Z}$ due to momentum-dependent electron-hole exchange interaction, which suppresses the Zeeman effect for finite momentum Bragg-umklapp-scattered excitons~\cite{smolenski2021,shimazaki2021}. While our spin-dependent calculations in Methods Sec.~13 confirm these expectations, they also reveal that exciton-electron interactions $V_\mathrm{X}$ can induce a Zeeman splitting for the umklapp peaks, whose magnitude increases with $|V_\mathrm{X}|$, and approaches that of the bare exciton for sufficiently strong interactions (see Fig.~\ref{fig:Fig3_Bfield}{\bf g}). The sign of this interaction-induced splitting is directly determined by the character of the interactions: it is the same (opposite) as that of the X Zeeman splitting for attractive (repulsive) $V_\mathrm{X}$. This is consistent with our experimental observations, confirming that WPs result from attractive particle–hole dressing of Bragg-scattered excitons by the WC, and further highlighting the central role of exciton–electron interactions in the investigated system. We note that, unlike WP$_\mathrm{T}$, the Zeeman splitting $\Delta_\mathrm{Z}(\mathrm{WP_S})$ of the singlet WP branch exhibits a complex, non-monotonic dependence on electron density. This behavior is likely linked to the spin–valley polarization degree of the electron system, which shows a local maximum where $\Delta_\mathrm{Z}(\mathrm{WP_S})$ reaches a minimum, suggesting that WP transitions could serve as sensitive probes of emergent spin phenomena in the electron crystal.

Our experiments demonstrate that optical excitations in WCs can not only sense electronic order but also strongly interact with it, becoming dressed into Wigner crystal polarons. The spectral properties of these quasiparticles are governed by collective excitations of the WC, rendering them powerful probes of the WC dynamics. This sensitivity may allow optical monitoring of the WC gap, that could be tuned at a fixed electron density by a proximal metallic screening layer~\cite{Stepanov2020,Liu2021c}, even at ultrafast timescales~\cite{Li2025b}. 

More broadly, the framework established in our work provides a route to exploring hybrid light–matter excitations in a wide range of strongly correlated electronic phases—including fractional Chern insulators~\cite{Park2023,zeng2023,lu2024}, kinetically driven magnetic orders~\cite{ciorciaro2023}, and superconductors~\cite{cao2018,xia2025,guo2025}—opening possibilities for optically accessing the fine structure of collective modes in strongly interacting quantum matter.

\subsection*{Acknowledgments}
We acknowledge fruitful discussions with Kai Klocke, Clemens Kuhlenkamp, and Ajit Srivastava. F.P. and M.K. acknowledge support from the Deutsche Forschungsgemeinschaft (DFG, German Research Foundation) under Germany’s Excellence Strategy–EXC–2111–390814868, the European Union (grant agreement No 101169765), as well as the Munich Quantum Valley, which is supported by the Bavarian state government with funds from the Hightech Agenda Bayern Plus. K.W. and T.T. acknowledge support from the JSPS KAKENHI (grant numbers 21H05233 and 23H02052), the CREST (JPMJCR24A5), JST and World Premier International Research Center Initiative (WPI), MEXT, Japan.

\subsection*{Author contributions}
T.S. conceived the experiments. L.W., F.M., and T.S. carried out the low-temperature magneto-optical measurements and analyzed the data. L.W. fabricated device A, while P.K. prepared device B. K.W. and T.T. grew the hBN crystals. F.P. performed theoretical modeling and numerical calculations under the guidance of M.K.. T.S., M.K., L.W., F.M., and F.P. wrote the manuscript. M.K. and T.S. supervised the project.

\section*{Methods}

\renewcommand{\figurename}{Extended Data Figure}
\renewcommand{\theHfigure}{M\arabic{figure}}
\setcounter{figure}{0}

\subsection*{1. Device fabrication}

The flakes used for device fabrication were mechanically exfoliated from bulk crystals (HQ Graphene WSe$_2$, NIMS hBN, NGS natural graphite) onto silicon substrates. They were stacked together inside an N$_2$-filled glove box using a standard dry-transfer method~\cite{Zomer2014}. In this technique, the flakes are aligned with submicron precision and subsequently picked up with a transparent polymer stamp consisting of a dome-shaped polydimethylsiloxane (PDMS) covered with a thin polycarbonate (PC) film. The finished vdW heterostructure was then released onto a silicon substrate with a 285-nm-thick SiO$_2$ layer and prepatterned electrodes (approximately 15~nm gold with about 5~nm titanium adhesion layer). Prior to the release step, the device was aligned with respect to the electrode pattern such that each few-layer graphene (FLG) flake (top gate, contact, and back gate) was connected to a separate electrode. Finally, the PC residue was removed by dissolving it in either chloroform or dichloromethane.

For both devices, the thicknesses of encapsulating hexagonal boron-nitride (hBN) layers were selected to ensure either constructive (device A) or destructive (device B) interference between resonant light reflected of various interfaces away from the WSe$_2$ monolayer region. This choice guarantees almost Lorentzian lineshapes of excitonic optical transitions in the WSe$_2$ monolayer reflectance contrast spectra, which greatly facilitates resolving weak spectral signatures of the Wigner crystal. In particular, device~A, along with two $\sim30$-nm-thick hBN layers separating WSe$_2$ from the top and bottom FLG gates, consisted of a third $\sim\SI{40}{\nano\metre}$-thick hBN capping layer (see Fig.~\ref{fig:Fig1}{\bf a}). In contrast, no hBN capping layer was used for device~B. For both samples, the doping density in the WSe$_2$ monolayer was tuned by applying top-gate-voltage $V_\mathrm{TG}$, while keeping the bottom gate at the ground potential. Device A showed no signatures of electrical hystrestis when $V_\mathrm{TG}$ was ramped in a loop. To avoid residual gate hysteresis for the device B (shown in Extended Data Figs.~\ref{fig:2nd_device_umklapp} and \ref{fig:2nd_device_melting}), its charge configuration was reset before each measurement by sweeping the gate voltage to \SI{-4}{\volt} with the imaging light on.

\subsection*{2. Experimental setup}

\begin{SCfigure*}[][t!]
	\includegraphics[scale = 1.05]{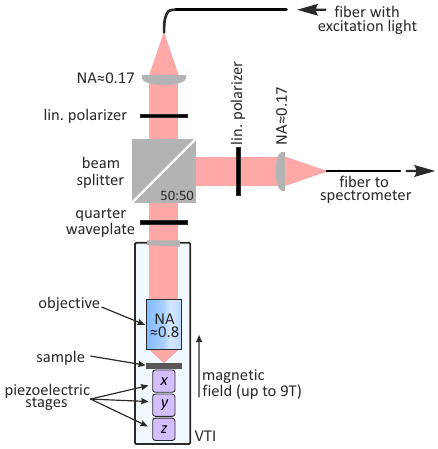}
	\caption{{\bf Experimental setup.} Simplified schematic of the experimental setup. \label{fig:setup}}
\end{SCfigure*}

Our experiments were carried out in a closed-cycle cryostat equipped with a superconducting magnet (up to \SI{9}{\tesla}). The sample was mounted on $x$–$y$–$z$ piezoelectric stages inside a variable-temperature insert (VTI) filled with He exchange gas to ensure efficient thermal contact, allowing the sample temperature to be tuned from \SI{1.6}{\kelvin} to room temperature. Reflectance measurements were performed in a confocal geometry (Extended Data Fig.~\ref{fig:setup}) using a supercontinuum laser (30 MHz repetition rate) coupled to a motorized filter and a single-mode fiber, enabling automatic selection of the excitation spectral range. The excitation light was collimated before entering the VTI through a wedged window, and then focused onto the sample surface to a diffraction-limited spot by a microscope objective with a large numerical aperture (NA = 0.8). The excitation power in front of the VTI window was monitored with a fast photodiode and actively stabilized using a PID controller coupled to a voltage-controlled variable optical attenuator. Except for the optical spin-orientation experiments, the power value was kept in the range of a few tens of \SI{}{\nano\watt} to ensure that the excitons did not perturb the probed electronic system. The light reflected from the sample was collected by the same objective, directed into the detection path by a beam splitter, coupled into a single-mode fiber, and finally dispersed by a spectrometer equipped with a Peltier-cooled CCD camera enabling recording of the spectra. The polarizations of both excitations and detection beams were controlled with a set of linear polarizers and achromatic waveplates.

\subsection*{3. Analysis of reflectance spectra}
\label{subsec:DRanalysis}

To determine $\Delta R/R_0$, the measured reflectance spectrum $R$ is normalized using the reference spectrum $R_{0}$ acquired in a region of the device featuring all layers except for the WSe$_2$ monolayer. The reflectance contrast is then determined at each photon energy as $R_{\mathrm{c}} \equiv \Delta R / R_0 = ((R - C_\mathrm{dark})- (R_0-C_\mathrm{dark})) / (R_0-C_\mathrm{dark})$, where $C_\mathrm{dark}$ denotes the dark counts of the CCD camera.

The derivative of the reflectance contrast with respect to the energy and gate voltage was obtained using a standard symmetric difference quotient method. Similarly as in our previous study~\cite{smolenski2021}, to reduce the noise, the numerical derivative is approximated as
$R_{\mathrm{c}}'(V_{\mathrm{G},n}) = \left[ (R_{\mathrm{c}}(V_{\mathrm{G},n+\delta}) - R_{\mathrm{c}}(V_{\mathrm{G},n-\delta}) \right]/ \left[ V_{\mathrm{G},n+\delta} - V_{\mathrm{G},n-\delta} \right]$ and $R_{\mathrm{c}}'(E_n) = \left[ (R_{\mathrm{c}}(E_{n+\delta'}) - R_{\mathrm{c}}(E_{n-\delta'}) \right] / \left[ E_{n+\delta'} - E_{n-\delta'} \right]$, where the index $n$ denotes subsequent points. The step $\delta$ in gate voltage and $\delta'$ in energy are chosen such that noise from abrupt changes of $R_{\mathrm{c}}$ occurring within $\Delta n_\mathrm{e} \lesssim \text{a few } 10^{10}\,\mathrm{cm}^{-2}$ or $\Delta E\lesssim 1$~meV is suppressed.

\subsection*{4. Fitting the spectral energies of optical resonances} \label{subsec:fitting}

\begin{SCfigure*}[][t!]
    \includegraphics{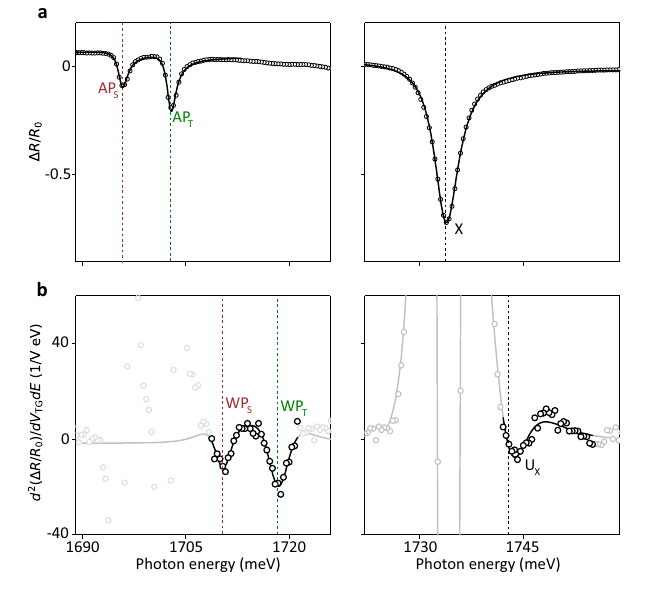}
	\caption{{\bf Fitting the spectral profiles of optical transitions.} ({\bf a,b}) Example reflectance contrast spectra ({\bf a}) and their second derivatives with respect to gate voltage and energy ({\bf b}) plotted in the spectral range of the AP/WP resonances (left) and X/U$\mathrm{X}$ resonances (right). The spectra were measured for device A at $T=1.6$~K, $B=0$, and $n_\mathrm{e}\approx3.6\cdot10^{11}\ \mathrm{cm}^{-2}$. Points mark the experimental data, while solid lines show the fitted lineshapes of optical resonances with the $\mathcal{L}$ or $\mathcal{D}$ spectral profiles described in Methods Sec.~4. In the case of umklapp transitions, the black (grey) curves/datapoints lie inside (outside) the spectral window in which the fit was performed. Vertical dashed lines mark the energies of optical transitions extracted from the fitted curves.}
	\label{fig:fitting}
\end{SCfigure*}

To determine the energies of the main excitonic transitions (X, AP$_\mathrm{S}$, and AP$_\mathrm{T}$) in the reflectance contrast spectra, we fit their spectral profiles with the following dispersive Lorentzian formula~\cite{Smolenski2019, shimazaki2020, smolenski2021}
\begin{equation}
\begin{split}
\mathcal{L}(E_0,A,\gamma,\varphi;E)&= \\ & A\cos\varphi\frac{\gamma/2}{(E-E_0)^{2}+\gamma^{2}/4} + \\ & A\sin\varphi\frac{E_0-E}{(E-E_0)^{2}+\gamma^{2}/4}+B(E),
\end{split}
\label{eq:disp_lorenzian}
\end{equation}
where $E$ denotes the photon energy, while $A$, $E_0$, and $\gamma$ represent, respectively, the amplitude, peak energy, and linewidth of the analyzed optical resonance. The parameter $\varphi$ represents the phase shift induced by interference of light reflected off different interfaces within the heterostructure, which depends on both the resonance energy and amplitude. Finally, $B(E)$ accounts for a background term stemming from spectral proximity of other resonances, which is assumed to be either constant $B(E)=C$ or polynomially dependent on the photon energy. For the APs, we assume that both transitions have a common phase. Moreover, at densities below $\sim10^{11}\ \mathrm{cm}^{-2}$---where the APs become very weak---their phase and linewidths are fixed at values obtained for a higher $n_\mathrm{e}$. Coherently with previous works~\cite{Smolenski2019, shimazaki2020, smolenski2021}, this method allows us to precisely reproduce the lineshapes of all three main resonances, as shown in Extended Data Fig.~\ref{fig:fitting}{\bf a}.

Owing to their markedly smaller intensities, the umklapp-scattered exciton and Wigner crystal polaron resonances cannot be fitted using the above procedure. Instead, to extract their spectral positions, we fit directly the differentiated $d^2(\Delta R/R_0)/dV_\mathrm{TG}dE$ spectra, where the umklapp transitions are greatly amplified due to their much larger blueshift as compared to the main excitonic resonances. To quantitatively describe the resulting spectral profiles, we assume that the evolution of U$_\mathrm{X}$ or WP lineshape $\mathcal{L}_\mathrm{U}(E_\mathrm{U},A_\mathrm{U},\gamma_\mathrm{U},\varphi_\mathrm{U};E)$ is dominated by the linear-in-density energy blueshift. This allows us to approximate its differentiated spectral profile $\mathcal{D}_\mathrm{U}(E_\mathrm{U},A_\mathrm{U},\gamma_\mathrm{U},\varphi_\mathrm{U};E)$ in the following way:
\begin{widetext}
\begin{equation}
    \begin{aligned}
\mathcal{D}_\mathrm{U}(E_\mathrm{U},A_\mathrm{U},\gamma_\mathrm{U},\varphi_\mathrm{U};E)&=\frac{d^2}{dV_\mathrm{TG}dE}\mathcal{L}_\mathrm{U}(E_\mathrm{U},A_\mathrm{U},\gamma_\mathrm{U},\varphi_\mathrm{U};E) \approx \left(\frac{d E_\mathrm{U}}{d \mathrm{V}_\mathrm{TG}}\right) \cdot \frac{\partial}{\partial E_\mathrm{U}} \frac{d}{d E} \mathcal{L}_\mathrm{U}(E_\mathrm{U},A_\mathrm{U},\gamma_\mathrm{U},\varphi_\mathrm{U};E)=\\
&=\tilde{A}_\mathrm{U}\frac{\cos\varphi_\mathrm{U}[\gamma_\mathrm{U}^2/4-3(E-E_\mathrm{U})^2]\gamma_\mathrm{U}/2+\sin\varphi_\mathrm{U}(E-E_\mathrm{U})[-3\gamma_\mathrm{U}^2/4+(E-E_\mathrm{U})^2]}{[\gamma_\mathrm{U}^2/4+(E-E_\mathrm{U})^2]^3},
\label{eq:Rc_fit}
\end{aligned}
\end{equation}
\end{widetext}
where $\tilde{A}_\mathrm{U}=2A_\mathrm{U}\cdot dE_\mathrm{U}/dV_\mathrm{TG}$ is the effective amplitude of the analyzed umklapp transition in the differentiated spectrum, while $\gamma_\mathrm{U}$, $\varphi_\mathrm{U}$, and $E_\mathrm{U}$ denote, respectively, its linewidth, phase, and energy. In view of their negligible amplitudes, the phases of all three umklapp resonances can be fixed at the same value corresponding to the low-oscillator-strength limit of the phase of an optical resonance in the analyzed device~\cite{smolenski2021}. It is determined based on the transfer matrix (TM) simulations of the device reflectivity spectrum~\cite{back2018,scuri2018} (see the next Methods section) and further confirmed by the phase of two AP resonances in the low-density limit, yielding $\varphi_\mathrm{U}\approx3.0$ for device A, and $\approx0.3$ for device B. To facilitate the fitting procedure, the linewidths of the umklapp resonances are always constrained within reasonable ranges, and assumed to increase with $n_\mathrm{e}$ for the U$_\mathrm{X}$. The fitting itself is carried out within a narrow spectral window around a given resonance. In the case of WPs, each transition is first individually fitted. The obtained parameters are then partially constrained, and used as a starting point for a final fit of the two WP branches with the sum of two $\mathcal{D}_\mathrm{U}$ profiles that is further corrected by a background term accounting for the tails of the main X and AP resonances in the spectrum. This is justified by their relatively large energy distance from both WP resonances. This assumption does not hold for the U$_\mathrm{X}$ resonance, which spectrally overlaps with the main exciton resonance. To account for this, the U$_\mathrm{X}$ is fitted using a sum of its expected profile $\mathcal{D}_\mathrm{U}$ and numerically-computed second derivative of the dispersive Lorentzian profile fitted beforehand to the X resonance. As shown in Extended Data Fig.~\ref{fig:fitting}{\bf b}, these procedures enable us to accurately reproduce the spectra of both WP and U$_\mathrm{X}$ transitions, and thereby quantitatively extract their energies.

\subsection*{5. Determination of the spectral weights of optical resonances} \label{subsec:fitting}

\begin{figure*}
    \includegraphics{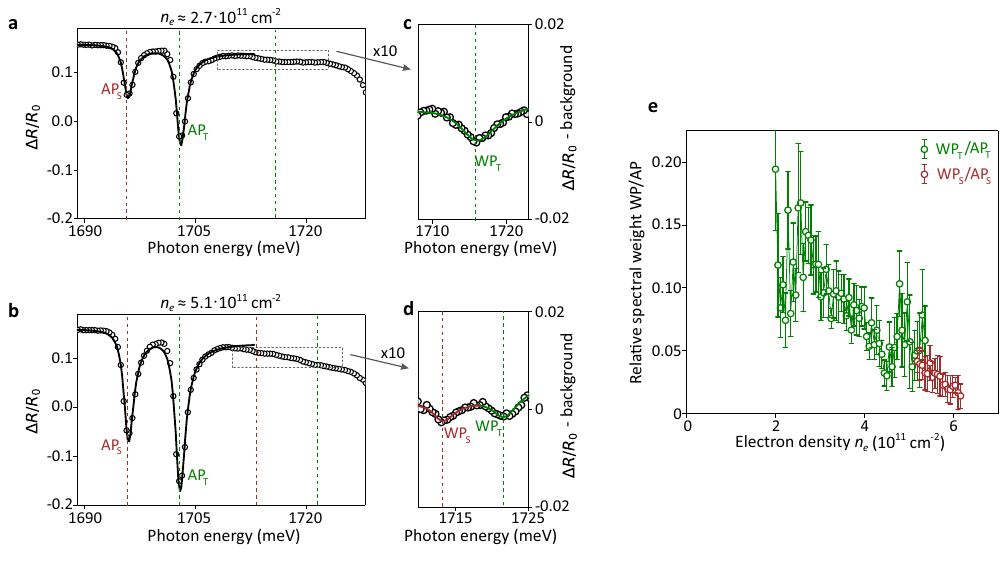}
	\caption{{\bf Oscillator strengths of optical resonances based on TM fitting.} ({\bf a,b}) Example reflectance contrast spectra measured for the main device at $T=1.6$~K, $B=0$, and two different electron densities (as indicated). The solid lines mark the fits of the AP spectral profiles with the TM model. ({\bf c,d}) Close-ups of the WP energy ranges (marked by dashed rectangles in {\bf a,b}), showing background-corrected spectra together with the corresponding TM fits of the WP spectral profiles. ({\bf e}) Relative oscillator strengths of WP$_\mathrm{T}$ and WP$_\mathrm{S}$ resonances divided by the spectral weights of corresponding main AP resonances as a function of electron density. Note that determination of WP intensities is possible only when they are spectrally well-separated from X/AP resonances.}
	\label{fig:TM_fits}
\end{figure*}

While dispersive Lorentzian fitting described in Methods Sec.~4 allows for precise determination of the spectral positions of optical resonances, the amplitudes obtained in such a way do not directly correspond to the oscillator strengths. Whenever we are interested in the latter (for normalizing the WC phase diagrams in Fig.~\ref{fig:Fig2_phase_diag}{\bf c} and Extended Data Fig.~\ref{fig:2nd_device_melting}{\bf c} as well as for analyzing the density-dependent AP intensities in Extended Data Figs.~\ref{fig:V0_cal} and~\ref{fig:dens_cal}), we employ a more complex transfer-matrix (TM) model~\cite{zeytinoglu2017,back2018,scuri2018,Smolenski2022}. In this case, each resonance is described with the following complex susceptibility $\chi(E)=-(\hbar c/E_\mathrm{0})\gamma_\mathrm{rad}/(E-E_0+i\gamma_\mathrm{nrad}/2)$, where $E$ is the photon energy, $E_0$ is the resonance energy, while $\gamma_\mathrm{nrad}/\hbar$ and $\gamma_\mathrm{rad}/\hbar$ denote, respectively, its nonradiative and free-space radiative decay rates, with the latter being directly proportional to the oscillator strength. Assuming the hBN thicknesses from Methods Sec.~1 and standard values of hBN/graphene/SiO$_2$ refractive indices~\cite{smolenski2021,Smolenski2022}, we can aptly reproduce the interference pattern within the vdW heterostructure, thus correctly capturing the lineshapes of optical resonances, as shown for example spectra in Extended Data Figs.~\ref{fig:TM_fits}{\bf a,b}. In these fits, analogously as for dispersive Lorentzian profiles, TM-calculated spectra are overlaid on a background term. Moreover, nonradiative broadening $\gamma_\mathrm{nrad}$ controlling the linewidths of APs is assumed to be the same for both singlet and triplet branches, and fixed in the low-density regime below a few $10^{10}\ \mathrm{cm}^{-2}$. By performing such fitting at various electron densities, we can directly determine $n_\mathrm{e}$-dependent oscillator strengths of the main AP/X resonances $f_\mathrm{AP/X}(n_\mathrm{e})=\gamma_\mathrm{rad,AP/X}(n_\mathrm{e})/\gamma_\mathrm{rad,X}$ relative to that of the X at charge neutrality (set by $\gamma_\mathrm{rad,X}$).

Given their inherently weak nature, WP transitions cannot be easily fitted with a similar TM model, particularly near the WC melting point. In this regime, tracing WP intensities requires alternative approaches (exploited in Methods Sec.~8) that do not allow for quantitative extraction of $f_\mathrm{WP}$. This is, however, possible at low temperatures in a selected density range where the WC is stable and WP transitions are spectrally well-separated from the X/AP resonances. To facilitate the TM fitting in this regime, we first subtract fitted AP and X spectral profiles, as well as similarly-corrected reflectance spectrum measured at charge-neutrality (to suppress the influence of residual etaloning in our setup). This data is then further corrected by subtracting linearly-dependent background fitted in the spectral range between X and AP$_\mathrm{T}$. Finally, the WP resonances in the resulting background-corrected spectrum are fitted using above-described TM model with fixed energies obtained based on dispersive Lorentzian model. As shown in Extended Data Figs.~\ref{fig:TM_fits}{\bf c,d}, this procedure reproduces the WP lineshapes reasonably well, allowing us to extract their density-dependent spectral weights $f_\mathrm{WP}/f_\mathrm{AP}$ relative to the corresponding AP branches (Extended Data Fig.~\ref{fig:TM_fits}{\bf e}). Interestingly, despite the fact that their peak-to-peak amplitudes are more than a few tens of times smaller than that of APs, the WPs exhibit sizably larger linewidth, presumably due to additional broadening channels stemming from the WC disorder that does not directly affect the APs themselves. For this reason, $f_\mathrm{WP}/f_\mathrm{AP}$ can be as high as 10--20\% in the low-density regime, supporting our conclusion of substantial hybridization of AP and WP transitions. Still, we stress that in view of complex background subtraction procedures, the extracted $f_\mathrm{WP}/f_\mathrm{AP}$ values are fraught with large uncertainties, which are likely larger than statistical error bars shown in Extended Data Fig.~\ref{fig:TM_fits}{\bf e}.

\subsection*{6. Calibration of the doping density} \label{subsec:density_calibration}
Since the energy splitting between the U$_\mathrm{X}$/WP transitions and the corresponding main X/AP resonances is determined by the WC lattice constant, which in turn depends on the electron density $n_\mathrm{e}$, it is crucial for our analysis to accurately determine $n_\mathrm{e}$. To this end, we describe our device as a parallel-plate capacitor, which allows us to express $n_\mathrm{e}$ as $n_\mathrm{e} = (C_\mathrm{TG}/e)\cdot(V_\mathrm{TG} - V_0)$, where $e$ is an elementary charge, $V_0$ is the voltage corresponding to the onset of electron filling into the conduction band, and $C_\mathrm{TG}$ is the geometrical capacitance between top gate and WSe$_2$ monolayer per unit area of the device.  

\begin{figure*}[]
    \includegraphics{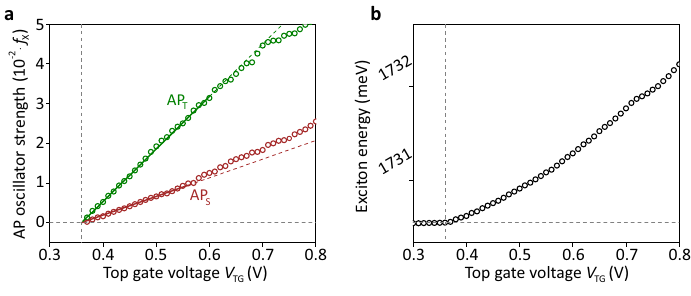}
    \caption{{\bf Determination of the doping onset voltage.} (\textbf{a}) Top-gate voltage dependent spectral weights $f_{\mathrm{AP_{S,T}}}$ of the AP resonances. The $f_{\mathrm{AP_{S,T}}}$ value are determined based on TM fits to the reflectance spectra, and expressed relative to oscillator strength $f_\mathrm{X}$ of the exciton at charge neutrality. The linear fits in the low-density regime (solid lines) are extrapolated to zero spectral weight to obtain $V_0$. (\textbf{b}) The X resonance energy determined based on dispersive Lorentzian fitting of the X spectral profile as a function of~$V_\mathrm{TG}$. The onset of the blueshift, marked by dashed lines, corresponds to $V_0$.}
    \label{fig:V0_cal}
\end{figure*}

To determine $V_0$, we analyze $V_\mathrm{TG}$-dependent AP oscillator strengths and X energies that are extracted based on TM or dispersive Lorentzian fitting of their spectral profiles (as described in Methods Secs.~4~and~5). Consistently with the Fermi polaron model~\cite{Sidler2017,Efimkin2017}, the spectral weights of singlet and triplet AP resonances exhibit linear increase with $V_\mathrm{TG}$ in the low-density regime, as shown in Extended Data Fig.~\ref{fig:V0_cal}{\bf a} presenting example results obtained at $T=1.6$~K. Importantly, linear fits to both increasing trends extrapolate to the zero oscillator strength at exactly the same voltage, which precisely corresponds to $V_0$. This value is consistent with that corresponding to the onset of doping-induced X blueshift (see Extended Data Fig.~\ref{fig:V0_cal}{\bf b}).

\begin{figure*}
\includegraphics{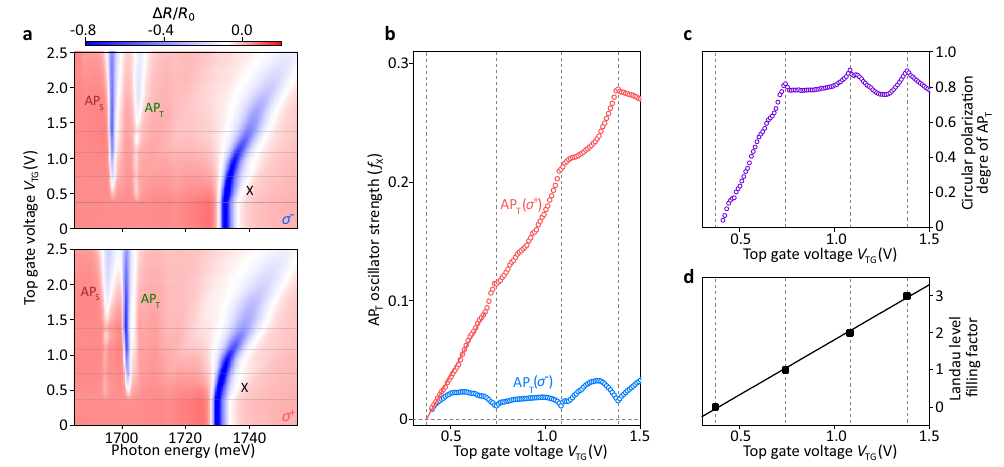}
	\caption{{\bf Electron density calibration based on the Shubnikov-de Haas oscillations in the optical spectra.} (\textbf{a}) $V_\mathrm{TG}$ evolutions of the reflectance contrast spectra acquired in $\sigma^-$ (top panel) and $\sigma^+$ (bottom panel) polarization at $T = 1.6$~K and $B = 8$~T. (\textbf{b}) Gate-voltage dependence of the AP$_\mathrm{T}(\sigma^\pm)$ oscillator strength extracted by fitting its spectral profile in $\sigma^\pm$ polarization with TM model. The value of $f_\mathrm{AP_T}(\sigma^\pm)$ is expressed relative to the exciton oscillator strength determined by fitting its spectral profile at charge neutrality (averaged over two circular polarizations). Solid line depicts the linear fit to the voltage-dependent $f_\mathrm{AP_T}(\sigma^+)$ in the low-density regime, which is used to obtain the $V_\mathrm{TG}$ corresponding to $\nu = 0$. (\textbf{c}) Circular polarization degree of the AP$_\mathrm{T}$ resonance determined as $\mathcal{P}_\mathrm{AP_T}=[f_\mathrm{AP_T}(\sigma^+)-f_\mathrm{AP_T}(\sigma^-)]/[f_\mathrm{AP_T}(\sigma^+)+f_\mathrm{AP_T}(\sigma^-)]$. (\textbf{d}) $V_\mathrm{TG}$ values corresponding to integer filling factors $\nu$. A solid line indicates linear fit with $\nu=(C_\mathrm{TG}/e)\cdot(h/eB)\cdot(V_\mathrm{TG}-V_0)$, whose slope is used to extract the value of $C_\mathrm{TG}/e$ allowing us to calibrate electron density $n_\mathrm{e}=(C_\mathrm{TG}/e) \cdot (V_\mathrm{TG}-V_0)$ based on the applied $V_\mathrm{TG}$. Grey dashed lines in all panels mark the voltages at which AP spectral weights exhibit cusps/dips, which correspond to the integer LL filling factors.
    \label{fig:dens_cal}}
\end{figure*}

The capacitance $C_\mathrm{TG}$ is, in turn, extracted based on Shubnikov–de Haas oscillations of the optical transitions at high magnetic field (\SI{8}{\tesla}). These oscillations arise from successive filling of LLs (Extended Data Fig.~\ref{fig:dens_cal}\textbf{a}), giving rise to characteristic cusps in energies and intensities of exciton-polaron transitions at integer LL filling factors~\cite{Smolenski2019, Liu2020, Smolenski2022}. In our analysis, we focus on low-density regime where the electrons fill primarily the states in a single valley, and trace local maxima (minima) of AP$_\mathrm{T}$ oscillator strength in $\sigma^+$ ($\sigma^-$) polarization (Figs~\ref{fig:dens_cal}\textbf{b},\textbf{c}). Together with $V_\mathrm{TG}$ corresponding to the onset of electron doping, this allows us to determine gate voltages corresponding to subsequent $\nu = 0, 1, 2, 3$. As shown in Extended Data Fig.~\ref{fig:dens_cal}\textbf{d}, they are very well reproduced by a linear relation $\nu=(C_\mathrm{TG}/e)\cdot(h/eB)\cdot(V_\mathrm{TG}-V_0)$, enabling us to extract $C_\mathrm{TG}/e = (5.9\pm0.3) \cdot 10^{11}\,\mathrm{cm^{-2}\,V^{-1}}$. Importantly, this result agrees well with the geometrical capacitance calculated based on the hBN thickness of \SI{30}{\nano\metre} and the hBN dielectric constant $\epsilon_\mathrm{hBN} = 3.1$~\cite{xu2021,Popert2022}.

\vspace{-0.3cm}
\subsection*{7. Calibration of sample temperatures}
\begin{figure*}[]
	\includegraphics[scale = 1]{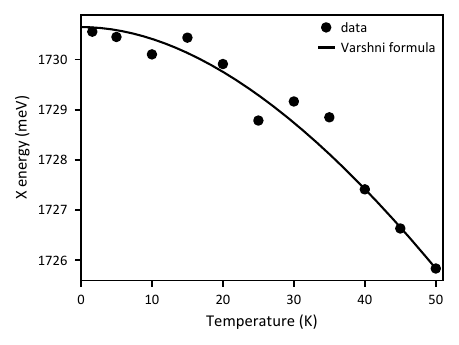}
	\caption{{\bf Sample temperature calibration.} Extracted X resonance energy at charge neutrality as a function of temperature, overlaid with the curve calculated using the Varshni formula.}
    \label{fig:T_cal}
\end{figure*}

\vspace{-0.1cm} \label{subsec:temp_calibration}
For all measurements, the sample temperature was determined using the temperature sensor of the VTI. To verify this calibration, we analyze the temperature-induced redshift of the exciton resonance, which arises from the shrinkage of the semiconducting bandgap $E_\mathrm{g}(T)$ and is well described by the Varshni formula~\cite{VARSHNI1967}:
\begin{equation}
E_\mathrm{g}(T) = E_0 - \frac{\alpha T^2}{T + \beta},
\end{equation}
where $E_0$ denotes the low-temperature limit, and $\alpha$ and $\beta$ are material-specific parameters previously reported as $\alpha = 0.424\ \mathrm{meV/K}$ and $\beta = \SI{170}{\kelvin}$ for monolayer WSe$_2$~\cite{arora2015}. As shown in Extended Data Fig.~\ref{fig:T_cal}, this dependence accurately reproduces the temperature evolution of the exciton energy at charge neutrality in our device, with $E_0$ serving as the only fitting parameter. The excellent agreement confirms the reliability of our temperature calibration.

\vspace{-0.3cm}
\subsection*{8. Analysis of the Wigner crystal melting}\vspace{-0.1cm}
To determine temperature- and doping-dependent intensities of umklapp and WP resonances, we pursue a similar approach to that introduced in our previous study~\cite{smolenski2021}. In the first step, $\mathrm{U}_\mathrm{X}$/WP spectral weights at a given $T$ and $n_\mathrm{e}$ are estimated by integrating out the absolute value of background-corrected differentiated reflectance signal $|d^2(\Delta R/R_0)/dV_\mathrm{TG}dE-B(E,n_\mathrm{e})|$ over a narrow window $E_\mathrm{min}<E<E_\mathrm{max}$ around each resonance. The background $B(E,n_\mathrm{e})$ is obtained separately for $\mathrm{U}_\mathrm{X}$/WP at each $n_\mathrm{e}$ by averaging polynomial profiles fitted to the data at high temperatures (above \SI{35}{\kelvin}), where the umklapp features become indiscernible. To facilitate quantitative comparison between different $\mathrm{U}_\mathrm{X}$/WP branches, the integration windows are taken to be equally wide for each of them. Moreover, their spectral positions are adjusted to compensate for doping- and temperature-induced shifts of umklapp transitions. In particular, at a given $n_\mathrm{e}$, the windows are displaced according to the temperature-induced energy shift of the neutral exciton $E_\mathrm{X}(T)$ at charge neutrality, as illustrated in Fig.~\ref{fig:Fig2_phase_diag}\textbf{b} and Extended Data Fig.~\ref{fig:2nd_device_melting}\textbf{b}.

In the second step, the above-determined spectral weights of umklapp resonances are normalized with the fitted oscillator strengths $f(n_\mathrm{e}, T)$ of the corresponding main X/AP resonances. This finally allows us to calculate relative intensities of exciton-umklapp or WP transitions as
\begin{equation}
    I_\mathrm{U}(n_\mathrm{e},T) = \frac{ \int_{E_\mathrm{min}}^{E_\mathrm{max}}\big| \tfrac{d^2 (\Delta R/R_0)}{dV_\mathrm{TG}\,dE} - B(E,n_\mathrm{e}) \big|dE}{f(n_\mathrm{e}, T)}.
\end{equation}
The extracted values of $I_\mathrm{U}(n_\mathrm{e},T)$ for U$_\mathrm{X}$ and both WP$_\mathrm{S,T}$ are plotted as a function of $n_\mathrm{e}$ and $T$ in Fig.~\ref{fig:Fig2_phase_diag}\textbf{c} and Extended Data Fig.~\ref{fig:2nd_device_melting}\textbf{c} for devices A and B, respectively, revealing the WC phase diagrams.

\vspace{-0.3cm}
\subsection*{9. Optical spin orientation of the Wigner crystal}\vspace{-0.1cm}

\begin{figure*}[]
	\includegraphics[scale = 1]{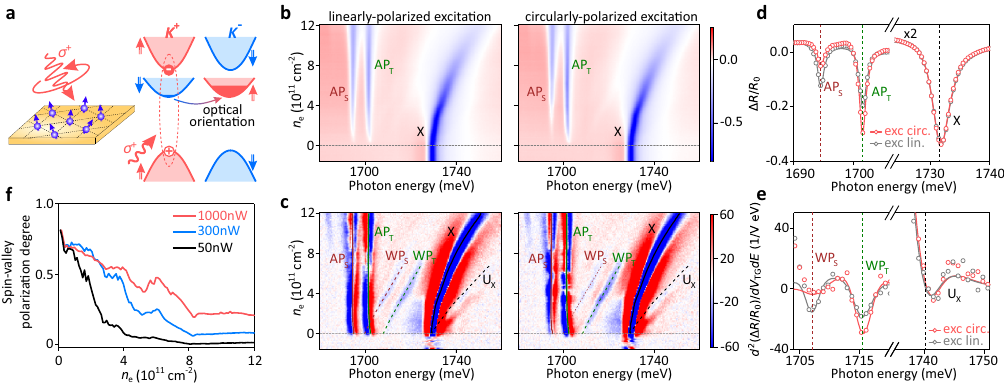}
	\caption{{\bf Optical spin orientation of the zero-field WC.} (\textbf{a}) Cartoon illustrating the idea of WC spin pumping under circularly-polarized optical excitation. (\textbf{b},\textbf{c}) Electron-density evolutions of bare (\textbf{b}) and differentiated (\textbf{c}) reflectance contrast spectra measured at $B=0$ and $T=1.6$~K under co-linearly (left) and co-circularly (right) polarized optical excitation with power $P=300$~nW and a broadband $\sim120$-meV-wide spectrum centered around the X resonance. To suppress the noise, spectrally-flat $n_\mathrm{e}$-dependent background was subtracted from the data in ({\bf b}). (\textbf{d},\textbf{e}) Linecuts through the maps in (\textbf{b},\textbf{c}) at $n_\mathrm{e}\approx3\cdot10^{11}\ \mathrm{cm}^{-2}$ showing the main resonances (\textbf{d}) and the corresponding WP/umklapp transitions (\textbf{e}) under two excitation polarization settings (same as Figs.~\ref{fig:Fig3_Bfield}{\bf i,j} from the main text). The data points in (\textbf{e}) were binned along the energy axis. Solid lines mark the fitted spectral profiles of the visible resonances. (\textbf{f}) Spin-valley polarization degree of the WC under circularly-polarized excitation of different powers (as indicated), determined as $\langle S_z\rangle=(f^\mathrm{circ}_\mathrm{AP_T}/f^\mathrm{lin}_\mathrm{AP_T}-f^\mathrm{circ}_\mathrm{AP_S}/f^\mathrm{lin}_\mathrm{AP_S})/(f^\mathrm{circ}_\mathrm{AP_T}/f^\mathrm{lin}_\mathrm{AP_T}+f^\mathrm{circ}_\mathrm{AP_S}/f^\mathrm{lin}_\mathrm{AP_S})$, where AP$^\mathrm{circ}_\mathrm{T,S}$ (AP$^\mathrm{lin}_\mathrm{T,S}$) denote oscillator strengths of the AP$_\mathrm{T,S}$ resonances obtained from TM fitting of their spectral profiles in circularly (linearly) excited reflectance spectra.}
    \label{fig:X_orient}
\end{figure*}

\begin{figure*}[]
    \includegraphics[width=13.5cm]{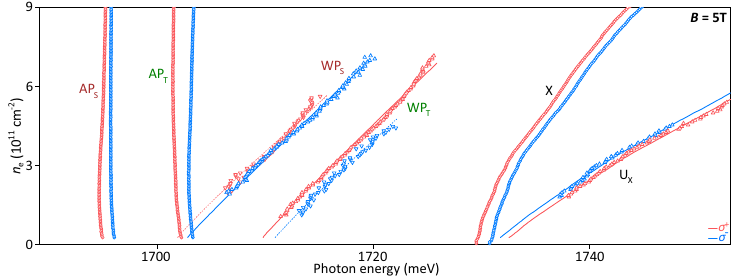}
	\caption{{\bf Energies of main and umklapp resonances at $B=5$~T.} Electron-density-dependent energies of both main and umklapp transitions determined at $B=5$~T in the two circular polarizations by fitting lineshapes of these resonances in bare or differentiated reflectance contrast spectra from Figs.~\ref{fig:Fig3_Bfield}{\bf b,c}. Solid lines mark the expected umklapp energies determined by fitting their splittings from the corresponding co-polarized main excitonic transitions as $h^2n_\mathrm{e}/\sqrt{3}m_\mathrm{X}+\Delta$ with fixed exciton mass $m_\mathrm{X}=0.68m_0$ and various offsets $\Delta$.}
	\label{ex:zeeman}
\end{figure*}

As described in the main text, the spins of the electrons forming a WC can be optically oriented even in the absence of a magnetic field. This is achieved by illuminating the sample with circularly polarized light (Fig.~\ref{fig:Fig3_Bfield}{\bf h} and Extended Data Fig.~\ref{fig:X_orient}{\bf a}), following an approach similar to that used in previous studies of charge-tunable WSe$_2$ monolayers~\cite{robert2021,goryca2021}. In contrast to these earlier works, which relied on non-resonant excitation, here we employ a broadband white light resonant with the main optical transitions in the WSe$_2$ monolayer spectrum. Extended Data Fig.~\ref{fig:X_orient}{\bf b} show density-dependent reflectance contrast spectra measured with such light of linear and circular polarization (using co-polarized detection). While in the former case the two AP$_\mathrm{T,S}$ resonances exhibit similar intensities, circular excitation drastically enhances their intensity difference in the low electron-density regime. Specifically, the AP$_\mathrm{T}$ resonance becomes considerably stronger, whereas the AP$_\mathrm{S}$ resonance is markedly weaker compared to the linearly excited spectra, as clearly seen in the linecut in Extended Data Fig.~\ref{fig:X_orient}{\bf d}. This observation provides direct evidence of optical spin orientation of the electrons, which are pumped out of the optically driven valley under circularly polarized excitation~\cite{robert2021}. Crucially, we find that this optical spin orientation does not compromise the long-range charge order of the WC, as directly revealed by the presence of exciton-umklapp resonances in the circularly-excited differentiated spectra (Extended Data Fig.~\ref{fig:X_orient}{\bf c}). Importantly, while the U$_\mathrm{X}$ transitions exhibit the same intensities in both linearly and circularly excited spectra (Extended Data Figs.~\ref{fig:X_orient}{\bf e}), the singlet and triplet branches of the WPs undergo the same changes as their AP counterparts, as highlighted in the main text.

Extended Data Fig.~\ref{fig:X_orient}{\bf f} shows the spin–valley polarization degree $\langle S_z\rangle$ of the optically oriented electron system as a function of electron density for different excitation powers. As expected for optical spin pumping, $\langle S_z\rangle$ at a given $n_\mathrm{e}$ typically increases with increasing excitation power. In parallel, higher $\langle S_z\rangle$ values are achieved at lower $n_\mathrm{e}$, presumably due to longer spin–valley relaxation times of the electrons in such a regime~\cite{goryca2021}.

\vspace{-0.3cm}
\subsection*{10. Data for the device B}
\begin{figure*}
	\includegraphics{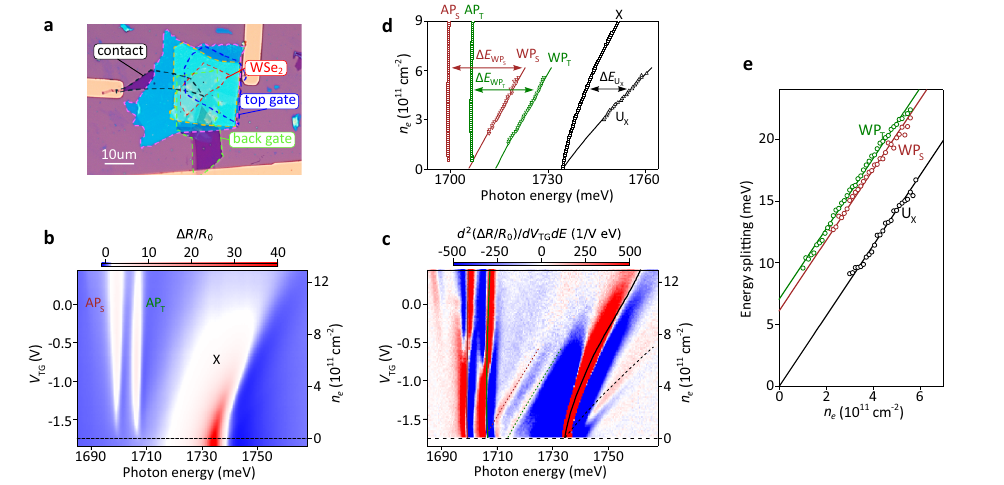}
	\caption{{\bf Wigner crystal polaron spectroscopy in  device B.} (\textbf{a}) Optical micrograph of device B. (\textbf{b, c}) Gate-voltage evolution of the reflectance contrast spectra (\textbf{b}) and their derivative with respect to gate voltage and photon energy (\textbf{c}) at $B = 0$ and $T = \SI{1.6}{\kelvin}$, with resonances marked in the same manner as for device A described in the main text. (\textbf{d}) Extracted energies of the X, AP$_\mathrm{T}$, AP$_\mathrm{S}$, U$_\mathrm{X}$, WP$_\mathrm{T}$, and WP$_\mathrm{S}$ resonances at low density. (\textbf{e}) Density evolution of the energy splittings between the main resonances and their corresponding umklapps. The dependencies are fitted using the same procedure as for the device A, yielding $m_\mathrm{X} \approx 0.61\,m_0$, $\Delta_{\mathrm{U_X}} = 0$, $\Delta_{\mathrm{WP_S}} = \SI{6}{\milli\electronvolt}$, and $\Delta_{\mathrm{WP_T}} = \SI{7}{\milli\electronvolt}$.} 
    \label{fig:2nd_device_umklapp}
\end{figure*}

\begin{figure*}
	\includegraphics[scale = 1]{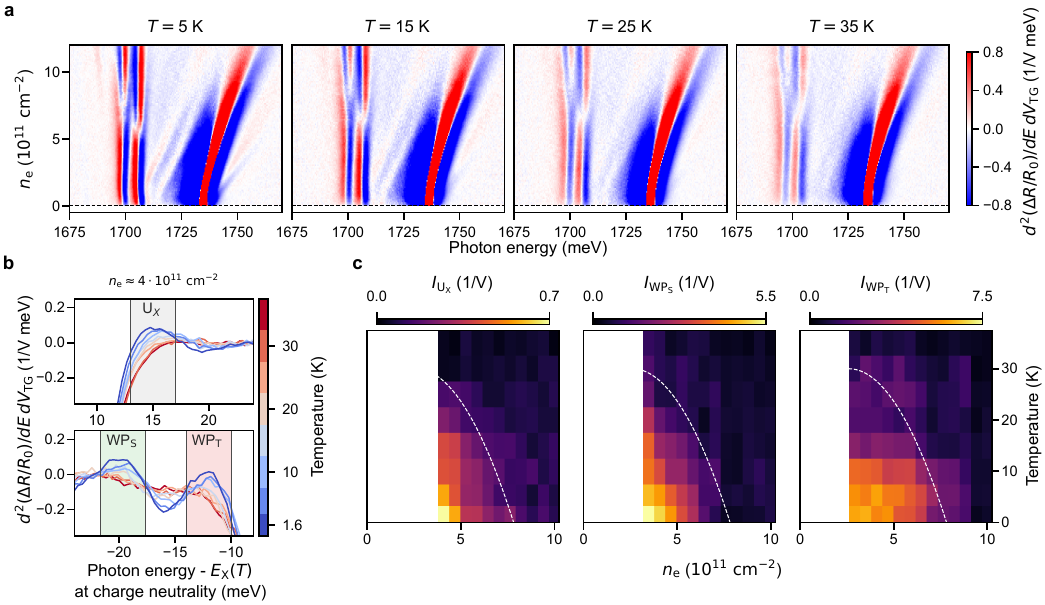}
	\caption{{\bf Temperature and density evolution of umklapp and WP resonances in the device B.} (\textbf{a}) Representative differentiated reflectance contrast spectra of device B at $T=\SI{5}{\kelvin}$, \SI{15}{\kelvin}, \SI{25}{\kelvin} and \SI{35}{\kelvin}. (\textbf{b}) Energy-aligned linecuts at $n_\mathrm{e}\approx4\cdot10^{11}\ \mathrm{cm}^{-2}$ through the differentiated spectra in \textbf{a}, taken at $T=\SI{1.6}{\kelvin}$--\SI{35}{\kelvin}. Shaded regions indicate the energy windows used to evaluate the spectral weights of the respective resonances. (\textbf{c}) Temperature--density phase diagrams of the Wigner crystal exctracted based on U$_\mathrm{X}$ (left), WP$_\mathrm{S}$ (middle) and WP$_\mathrm{T}$ (right) using the same procedure as in the main device.}
    \label{fig:2nd_device_melting}
\end{figure*}

\vspace{-0.1cm}
The results presented in the main text were reproduced in another sample (device B), which was fabricated without a top capping hBN layer (Extended Data Fig.~\ref{fig:2nd_device_umklapp}{\bf a}). In this device, the thicknesses of the top and bottom hBN encapsulation layers were chosen to ensure destructive interference between the resonant light reflected various interfaces away from the WSe$_2$ monolayer. For this reason, the optical resonances in the reflectance-contrast spectra of device B, while still exhibiting a close-to-Lorentzian lineshape, appear as peaks rather than dips, in contrast to the device A shown in the main text.

Low-temperature spectroscopy measurements of device B are summarized in Extended Data Figs.~\ref{fig:2nd_device_umklapp}{\bf b-e}. Consistent with the main text, we observe prominent signatures of exciton-umklapp transitions as well as both singlet and triplet WP resonances in the low-density regime, where the electrons form a WC. The energy splittings between each of these resonances and their main counterparts (i.e., X or AP$_\mathrm{S}$/AP$_\mathrm{T}$) exhibit the same behavior as for the device A. In particular, their increase with $n_\mathrm{e}$ can be described by linear relations of the same slope but with finite offsets in the case of WPs.

We also investigated thermal melting of the WC in device B. Extended Data Fig.~\ref{fig:2nd_device_melting} shows the corresponding temperature-dependent results, presented in a similar fashion as in Fig.~\ref{fig:Fig2_phase_diag} of the main text. Both U$_\mathrm{X}$ and WP$_\mathrm{S,T}$ resonances reveal similar dependence on $T$ and $n_\mathrm{e}$, exhibiting a characteristic dome-shaped phase diagram associated with the WC melting.

\vspace{-0.3cm}
\subsection*{11. Theory: Polaron formation}\vspace{-0.1cm}

To understand the formation of the Wigner crystal polaron, we consider the following model system:
\begin{equation}
    \begin{aligned}
        &H_\mathrm{X} =  \displaystyle\sum_{\mathbf{k}} \epsilon^\mathrm{X}_{\mathbf{k}} x^\dagger_{\mathbf{k}} x_{\mathbf{k}}, \\
        &H_\mathrm{e} = \displaystyle\sum_{\sigma \mathbf{k}} \epsilon^\mathrm{e}_{\mathbf{k}} c^\dagger_{\sigma \mathbf{k}} c_{\sigma \mathbf{k}} + \frac{1}{N}\displaystyle\sum_{\substack{\mathbf{k}, \mathbf{k}', \mathbf{q} \\ \sigma \sigma'}} V_{\mathbf{q}}^\mathrm{e-e} c^\dagger_{\sigma \mathbf{k}+\mathbf{q}}c^\dagger_{\sigma' \mathbf{k}'-\mathbf{q}} c_{\sigma' \mathbf{k}'}c_{\sigma \mathbf{k}}, \\
        &H_\mathrm{X-e} = \frac{1}{N}\displaystyle\sum_{\substack{\mathbf{k}, \mathbf{k}', \mathbf{q} \\ \sigma}} U^\mathrm{X-e}_\sigma c^\dagger_{\sigma \mathbf{k}+\mathbf{q}} c_{\sigma \mathbf{k}} x^\dagger_{\mathbf{k}'-\mathbf{q}} x_{ \mathbf{k}'},
    \end{aligned}
\end{equation}
with $\epsilon^\mathrm{X}_\mathbf{k} = \hbar^2 |\mathbf{k}|^2/(2 m_\mathrm{X})$ and $\epsilon^\mathrm{e}_\mathbf{k} = \hbar^2 |\mathbf{k}|^2/(2 m^*_\mathrm{b, e})$. The electron-exciton interaction $U^{\mathrm{X}-\mathrm{e}}_\sigma$ is taken to be attractive and contact-like. We treat the exciton as a tightly-bound impurity, which is immersed into an electronic state. The two electron flavors $\sigma \in \{K^\pm\}$ correspond to the bottom conduction band in the $K^\pm$ valley. Under the assumption that repulsive Coulomb interactions $V_{\mathbf{q}}^{\mathrm{e}-\mathrm{e}}$ stabilize the WC, we perform a mean-field decoupling of the electron interactions. The resulting mean-field Hamiltonian for the electrons consequently includes a periodic potential $V^\mathrm{e}(\mathbf{r}) \propto \langle n_\mathrm{e}(\mathbf{r})\rangle$, whose strength scales linearly with the electron density. Concretely, we parameterize the strength of the periodic potential seen by the electrons as $V^\mathrm{e} = \Delta_\mathrm{WC}$, where the WC gap $\Delta_\mathrm{WC}$ depends on the strength of electron-electron interactions. Since $V^\mathrm{e}$ scales linearly with density $n_\mathrm{e}$, we take the ratio $\Delta_\mathrm{WC}/E_\mathrm{F}$ to be constant in density, with the Fermi energy $E_\mathrm{F} = \pi \hbar^2 n_\mathrm{e}/m_\mathrm{b, e}^*$. We use $\Delta_\mathrm{WC}$ as a tuning parameter in our calculation. The triangular lattice constant of the WC is also related to the electron density $a_\mathrm{WC}^2 = 2/(\sqrt{3} n_\mathrm{e})$. We solve the electronic mean-field Hamiltonian by a change to its eigenbasis
\begin{equation}
H_\mathrm{e} = \displaystyle\sum_{{\sigma,\mathbf{k}, \lambda}}\varepsilon^\lambda_\mathbf{k} \gamma^\dagger_{\sigma \lambda \mathbf{k}} \gamma_{\sigma \lambda \mathbf{k}} \quad \text{with} \quad
    \gamma^\dagger_{\sigma \lambda \mathbf{k}} = \displaystyle\sum_{\mathbf{G}} \mathcal{U}_{\lambda \mathbf{G}}(\mathbf{k}) c^\dagger_{\sigma \mathbf{k}+\mathbf{G}},
\end{equation}
where $\mathbf{k}$ is now defined within the first Brillouin zone of the Wigner crystal, $\mathbf{G}$ are its reciprocal lattice vectors, and $\lambda$ are mean-field band indices. In this basis, the electron-exciton interactions become
\begin{equation}
    \begin{aligned}
        H_\mathrm{X-e} = \frac{1}{N} \displaystyle\sum_{\lambda \mu, \sigma}\displaystyle\sum_{\mathbf{G}, \mathbf{G}'} \displaystyle\sum_{\mathbf{k}, \mathbf{k}', \mathbf{q} \in BZ } &[U^\mathrm{X-e}_\sigma \Lambda_{\mathbf{G} \mathbf{G}'}^{\lambda \mu} \times \\ &\gamma^\dagger_{\sigma \mu \mathbf{k}} \gamma_{\sigma \lambda \mathbf{p}} x^\dagger_{\mathbf{q}-\mathbf{k}+\mathbf{p}+\mathbf{G}} x_{ \mathbf{q}+\mathbf{G}'}],
    \end{aligned}
\end{equation}
where the form factors
\begin{equation}
    \Lambda_{\mathbf{G} \mathbf{G}'}^{\lambda \mu} (\mathbf{k}, \mathbf{k}') = \frac{1}{N} \displaystyle\sum_{\mathbf{Q}} \mathcal{U}^*_{\lambda \mathbf{Q}-\mathbf{G}}(\mathbf{k})  \mathcal{U}_{\mu \mathbf{Q}-\mathbf{G}'}(\mathbf{k}') ,
\end{equation}
mediate interactions between excitons and the WC quasiparticles $\gamma^\dagger$. In particular, these exciton-electron interactions allow for scattering processes of excitons with different reciprocal lattice vectors $\mathbf{G}$.
To describe the formation of the WC polaron, we employ the Chevy approximation~\cite{chevyUniversalPhaseDiagram2006}, which was previously used to describe polaron formation in the presence of charge ordering~\cite{amelioPolaronFormationInsulators2024, pichler2025purelyelectronicmodelexcitonpolaron} and has been shown to yield consistent results with exact diagonalization~\cite{amelioPolaronSpectroscopyInteracting2024}. Concretely, we make a variational ansatz for the many-body wave function
\begin{equation}
\begin{aligned}
        | \Psi_\mathbf{p} \rangle = &\displaystyle\sum_{\mathbf{G}} \varphi_\mathbf{G}(\mathbf{p}) x^\dagger_{\mathbf{p}+\mathbf{G}} | \mathrm{GS} \rangle \\ +& \displaystyle\sum_{\mathbf{G}, \sigma} \displaystyle\sum_{\mathbf{k}, \mathbf{q}, \alpha, \beta} \phi_{\mathbf{G}, \sigma}^{\mathbf{k} \mathbf{q} \alpha \beta} x^\dagger_{\mathbf{p}+\mathbf{q}-\mathbf{k}+\mathbf{G}} \gamma^\dagger_{\sigma \alpha \mathbf{k}} \gamma_{\sigma \beta \mathbf{q}} | \mathrm{GS} \rangle \label{eq:Chevy}
\end{aligned}
\end{equation}
restricted to the subspace of the exciton interacting with a single particle-hole excitation of the WC. Here $| \mathrm{GS} \rangle$ is the electronic ground state.
By  projecting the full Hamiltonian onto the Chevy-ansatz, we find the following equations:
\begin{widetext}
\begin{equation}
    \begin{aligned}
        &\displaystyle\sum_{\mathbf{G}'}\big[\epsilon^\mathrm{X}_{\mathbf{p}+\mathbf{G}} \delta_{\mathbf{G}, \mathbf{G}'} +V^\mathrm{X}_{\mathbf{G}-\mathbf{G}'} +\displaystyle\sum_{\mathbf{q} \beta, \sigma} U^{\mathrm{X}-\mathrm{e}}_\sigma \Lambda_{\mathbf{G} \mathbf{G}'}^{\beta \beta}(\mathbf{q}, \mathbf{q} ) \big]\varphi_{\mathbf{G}'}(\mathbf{p})  +\displaystyle\sum_{\mathbf{G}' \sigma}\displaystyle\sum_{\mathbf{q} \beta,  \mathbf{k} \alpha} U^{\mathrm{X}-\mathrm{e}}_\sigma \Lambda_{\mathbf{G} \mathbf{G}'}^{\beta \alpha}(\mathbf{q}, \mathbf{k} ) \phi_{\mathbf{G'}, \sigma}^{\mathbf{k} \mathbf{q} \alpha \beta}(\mathbf{p}) = E_\mathbf{p} \varphi_{\mathbf{G}}(\mathbf{p}) 
    \end{aligned} \label{eq:Cheveq1}
\end{equation}
and
\begin{equation}
    \begin{aligned}
        &\displaystyle\sum_{\mathbf{G}'}\big[(\epsilon^\mathrm{X}_{\mathbf{p}+\mathbf{q}-\mathbf{k}+\mathbf{G}} + \varepsilon^\alpha_{\mathbf{k}} - \varepsilon^\beta_{\mathbf{q}}) \delta_{\mathbf{G}, \mathbf{G}'} + V^\mathrm{X}_{\mathbf{G}-\mathbf{G}'} + \displaystyle\sum_{\mathbf{q}' \beta', \sigma} U^{\mathrm{X}-\mathrm{e}}_\sigma \Lambda_{\mathbf{G} \mathbf{G}'}^{\beta' \beta'}(\mathbf{q}', \mathbf{q}' )\big]\phi_{\mathbf{G}', \sigma}^{\mathbf{k} \mathbf{q} \alpha \beta}(\mathbf{p}) +\displaystyle\sum_{\mathbf{G}' \sigma} U^{\mathrm{X}-\mathrm{e}}_\sigma \Lambda_{\mathbf{G} \mathbf{G}'}^{\alpha \beta}(\mathbf{k}, \mathbf{q} )\varphi_{\mathbf{G}'}(\mathbf{p}) + \\ + &\displaystyle\sum_{\mathbf{G}' \mathbf{k}' \mathbf{\alpha}'} U^{\mathrm{X}-\mathrm{e}}_\sigma \Lambda_{\mathbf{G} \mathbf{G}'}^{\alpha \alpha'}(\mathbf{k}, \mathbf{k}' ) \phi_{\mathbf{G}', \sigma}^{\mathbf{k}' \mathbf{q} \alpha' \beta}(\mathbf{p}) -  \displaystyle\sum_{\mathbf{G}' \mathbf{q}' \mathbf{\beta}'} U^{\mathrm{X}-\mathrm{e}}_\sigma \Lambda_{\mathbf{G} \mathbf{G}'}^{\beta' \beta}(\mathbf{q}', \mathbf{q} ) \phi_{\mathbf{G}', \sigma}^{\mathbf{k} \mathbf{q}' \alpha \beta'}(\mathbf{p})  = E_\mathbf{p} \phi_{\mathbf{G}, \sigma}^{\mathbf{k} \mathbf{q} \alpha \beta}(\mathbf{p}),
    \end{aligned} \label{eq:Cheveq2}
\end{equation}
\end{widetext}
which we solve numerically to obtain the wave function $| \Psi_\mathbf{p} \rangle$. Here, we have also explicitly included a periodic potential $V^\mathrm{X}$ for the excitons, which they experience due to the formation of the WC. 
For the Chevy calculations used to generate Fig.~\ref{fig:Fig_theory}\textbf{c}-\textbf{f} we include $N_\mathrm{bands} = 3$ bands, and $N_\mathbf{G} = 3\times 3$ reciprocal lattice vectors, centered around $\mathbf{G}=0$. The momentum sums over the first Brillouin zone contain $N=16$ momenta. We use an exciton mass $m_\mathrm{X} = 0.7 m_0$ and an electron mass $m_\mathrm{b, e}^* = 0.45 m_0$~\cite{Kormanyos2015, christianen2025asymmetrictrionsmonolayertransition}. We have checked that our qualitative conclusions do not depend on the precise value of these microscopic parameters. 
For the results presented in Fig.~\ref{fig:Fig_theory}\textbf{c}-\textbf{e}, we fix the electron density to $n_\mathrm{e} = 5\:\cdot\: 10^{11}\;\mathrm{cm}^{-2}$. We determine the electron-exciton interaction strength $|U^{\mathrm{X}-\mathrm{e}}|$ based on the splitting between AP and X, which we fix to $33$ meV, roughly corresponding to the experimental value for the singlet AP$_\mathrm{S}$ in the zero density limit. 
From bottom panel to top panel in Figs.~\ref{fig:Fig_theory}\textbf{c}-\textbf{e}, the electron and exciton potential are $V^\mathrm{e}=0$,  $V^\mathrm{X}=0$ (\textbf{e}), $V^\mathrm{e}=0.01E_\mathrm{F}$, $V^\mathrm{X}=0.3E_\mathrm{F}$ (\textbf{d}), $V_\mathrm{e} = 1.80E_\mathrm{F}$, $V^\mathrm{X}=0.3E_\mathrm{F}$ (\textbf{c}). The respective values are representative examples for the three regimes of Wigner crystal strength discussed in the main text. The three panels shown in Figs.~\ref{fig:Fig_theory}\textbf{c}-\textbf{e} should be understood as illustrative examples in the three regimes and not as quantitative predictions. 
For all periodic potentials, we assume a simple harmonic form, where $V_\mathbf{G}=\mathrm{const.}$ for the first shell of reciprocal lattice vectors and zero otherwise.

To make comparisons with the optically measured spectra, we calculate the zero-momentum exciton spectral function 
\begin{equation}
    \mathcal{A}_\mathrm{X}(\mathbf{p}=0, \omega) = \displaystyle\sum_n |\langle \Psi_{n,\mathbf{p}} |x^\dagger_{\mathbf{p}}|0\rangle|^2 \delta(\omega - E_{n,\mathbf{p}}),
\end{equation}
where the sum runs over all eigenstates $n$ of Eqs.~\eqref{eq:Cheveq1},~\eqref{eq:Cheveq2}. To simplify the analysis of the theoretical model, we assume spin-independent electron-exciton interactions, which result in a single AP resonance. As expected, we find that the splitting between the exciton and its umklapp peak predominantly depends on the exciton mass $m_\mathrm{X}$ and the electron density $n_\mathrm{e}$, and is not sensitive to the WC gap $\Delta_\mathrm{WC}$ and the electron-exciton interaction $U^{\mathrm{X}-\mathrm{e}}$. By contrast, the splitting between the AP the WP sensitively depends on both the WC gap $\Delta_\mathrm{WC}$ and the electron-exciton interaction $U^{\mathrm{X}-\mathrm{e}}$. Furthermore, the oscillator strength of the exciton umklapp peak depends on the strength of the potential experienced by the excitons $V^\mathrm{X}$, while the oscillator strength of the WP predominantly depends on $V^\mathrm{e}$, implying that dressing due to electron-exciton interactions plays a dominant role.

\subsection*{12. Theory: Hybridization model}
The experimentally observed splitting of the Wigner crystal polaron $\Delta E_\mathrm{WP}$ from the AP can be explained with a simple hybridization model, where the electron-exciton interactions induce a hybridization between the WP and AP transitions, see Eq. (1) of the main text, reproduced here:
\begin{equation*}
    H_\mathrm{hyb} = \begin{pmatrix}
        E_\mathrm{AP} && V_\text{hyb}(n_\mathrm{e}) \\
        V^*_\text{hyb}(n_\mathrm{e}) && E_\mathrm{AP} + \frac{h^2}{\sqrt{3}} \frac{n_\mathrm{e}}{m_\mathrm{X}}
    \end{pmatrix}.
\end{equation*}
The hybridization strength $V_\text{hyb}$ depends both on the effective contact interaction $U^\mathrm{X-e}_\text{eff}$ and the WC gap $\Delta_\mathrm{WC}$, which enters in the electron-exciton interactions through the form factors $\Lambda^{\mu \nu}_{\mathbf{G}, \mathbf{G}'}$. 
Keeping the number of mean-field bands included in our calculation fixed implies that the form factors are density independent, so that the only density dependence of the hybridization strength $V_\text{hyb}$ stems from $U^{\mathrm{X}-\mathrm{e}}_\text{eff}$. The effective interactions $U^\mathrm{X-e}_\mathrm{eff}$ can be estimated by the T-matrix that has a logarithmic divergence at low densities in two-dimensional systems $\propto 1/\log(n_0/n_\mathrm{e})$~\cite{Kuhlenkamp2022}, where $n_0 = m^*_\mathrm{b,e} E_\mathrm{b}/(\pi\hbar^2)$ is the typical density associated with the trion binding energy $E_\mathrm{b}$.
Within this hybridization model, the splitting between the AP and the WP is given by
\begin{equation}
    \Delta E_\text{WP} = \sqrt{\left(\frac{h^2}{\sqrt{3}} \frac{n_\mathrm{e}}{m_\mathrm{X}}\right)^2 + 4 |V_\text{hyb}(n_\mathrm{e})|^2}.
\end{equation}
Parameterizing the hybridization as $|V_\text{hyb}(n_\mathrm{e})|^2 = |V_\text{hyb}^0 / \log(n_0/n_\mathrm{e})|^2$, describes the density dependence of the splittings $\Delta E_\mathrm{WP_\mathrm{S, T}}$ for both Wigner crystal polarons with a single parameter $V_\text{hyb}^0$. We use the data for $\mathrm{WP}_\mathrm{S}$ to extract $V_\mathrm{hyb}^0 = 15.82 \; \mathrm{meV}$. This value is then used to predict the density dependence of $\mathrm{WP}_\mathrm{T}$, in good agreement with the experimental data (see Fig.~\ref{fig:Fig_theory}\textbf{g}). To calculate $n_0 = m^*_\mathrm{b,e} E_\mathrm{b}/(\pi\hbar^2)$, the respective experimental values of zero-density binding energies of $\mathrm{AP}_\mathrm{S}$ and $\mathrm{AP}_\mathrm{T}$ are used. Note that in our model we assume $\Delta_{\mathrm{WC}} / E_\mathrm{F}$ to be independent of the electron density, which is a good approximation deep within the WC phase. As the density approaches the critical value $n_c$, where the WC melts, $\Delta_{\mathrm{WC}} / E_\mathrm{F}$ decreases. This additional density dependence of $V_\text{hyb}(n_\mathrm{e})$, arising from the quantum melting of the WC, is not captured in our simple model. Nevertheless, even at high densities, the predictions of the model remain in good agreement with the experimental data. This is because, in this regime, the dominant contribution to the splitting $\Delta E_{\mathrm{WC}}$ originates from the diagonal Bragg-scattering term $\sim n_\mathrm{e} / m_\mathrm{X}$, which is independent of $\Delta_{\mathrm{WC}}$.

\subsection*{13. Theory: Magnetic field dependence}
To model the Zeeman splittings of the optical resonances involving umklapp-scattered excitons, we follow the approach of Ref.~\cite{shimazaki2021, salvador2022}. Concretely, we calculate the magnetic-field-induced splitting of an umklapp-scattered exciton as a function of the sign and strength of the electron-exciton interactions. Due to electron-hole exchange interactions, the $K^+$ and $K^-$ valley excitons (which at zero momentum couple to $\sigma^+$ and $\sigma^-$ polarized light, respectively) are coupled at finite momentum~\cite{ZhuStraintuning2013, yuDiracConesDirac2014, shimazaki2021}. In the spin-valley basis $\mathbf{x}^\dagger_\mathbf{k} = (x^\dagger_{ \mathbf{k}, +}, \; x^\dagger_{\mathbf{k}, -})$, the excitons are described by the Hamiltonian $H=H_0 + H_\mathrm{pot} + H_\mathrm{z}$, with
\begin{equation}
    \begin{aligned}
        &H_0 = \displaystyle\sum_\mathbf{k} \mathbf{x}^\dagger_\mathbf{k} \left[ \frac{\hbar^2\mathbf{k}^2}{2 m_\mathrm{X}} +J\frac{|\mathbf{k}|}{K} \begin{pmatrix}
            1 && e^{-2i\theta_\mathbf{k}} \\
            e^{2i\theta_\mathbf{k}} && 1
        \end{pmatrix}\right]\mathbf{x}_\mathbf{k}, \\
        &H_\mathrm{pot} = \displaystyle\sum_{\mathbf{k}\in \mathrm{BZ}} \displaystyle\sum_{\mathbf{G}, \mathbf{G}'}\mathbf{x}^\dagger_{\mathbf{k}+\mathbf{G}} V^\mathrm{X}_{\mathbf{G}-\mathbf{G}'} \mathbf{x}_{\mathbf{k}+\mathbf{G}'}, \\
        &H_\mathrm{z} = \frac{g \mu_B B}{2} \displaystyle\sum_\mathbf{k} \mathbf{x}^\dagger_\mathbf{k} \sigma^z \mathbf{x}_\mathbf{k} .
    \end{aligned}
\end{equation}
The first term $H_0$ describes the kinetic energy of the excitons and their coupling due to electron-hole exchange interactions, parameterized by $J$. Here, $K=4\pi/(3a_{0})$ is the momentum of the $K^\pm$ valley, $a_0 \approx 0.3\; \mathrm{nm}$ is the lattice constant of WSe$_2$, and $\theta_\mathbf{k} = \arctan(k_y/k_x)$. The second term $H_\mathrm{pot}$ models the periodic potential felt by the excitons due to the presence of a Wigner crystal, which in this approximation is treated as a static background. The potential $V^\mathrm{X}$ originates from a mean-field decoupling of electron-exciton interactions. Consequently, a positive (negative) potential $V^\mathrm{X}$ corresponds to repulsive (attractive) interactions. Finally, $H_\mathrm{z}$ captures the bare Zeeman splitting due to a finite magnetic field $B$. We assume a $g$-factor of $g=4$. 

\begin{figure*}[]
	\includegraphics[scale = 0.5]{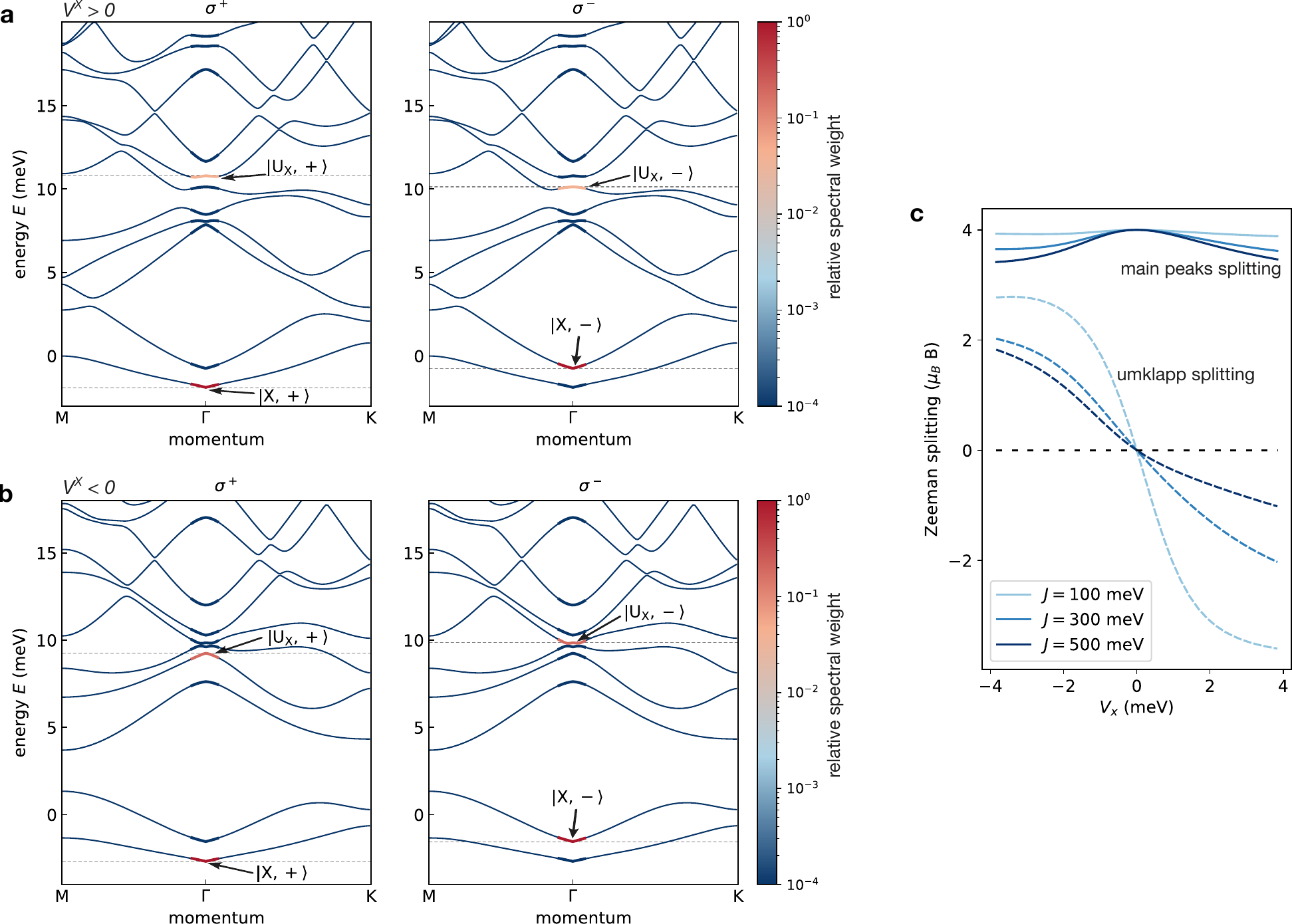}
	\caption{{\bf Interaction-dependent Zeeman splitting of optical resonances involving exciton-umklapp scattering.} Exciton bands for repulsive (\textbf{a}) and attractive (\textbf{b}) electron-exciton interactions in the presence of a periodically-ordered WC. The spectral weight of each state (relative to the strongest peak) is indicated through the color scale. Left (Right): coupling to $\sigma^-$ ($\sigma^+$) polarized light. We used $J=150$ meV and $|V^\mathrm{X}|=2$ meV at $B=5$T and $n_\mathrm{e} = 4\cdot 10^{11}$ cm$^{-2}$. (\textbf{c}) Zeeman splitting between the main peaks $|\mathrm{X}, \pm\rangle$ (solid lines) and between the umklapp peaks $|\mathrm{U}_\mathrm{X}, \pm\rangle$ (dashed lines) for various $J$ at $B=5$T as a function of the exciton potential strength $V^\mathrm{X}$. Negative $V^\mathrm{X}$, corresponding to attractive electron-exciton interactions, lead to positive Zeeman splitting for the umklapp peaks. Conversely, positive $V^\mathrm{X}$ (repulsive electron-exciton interactions) imply negative Zeeman splitting for the umklapp peaks.
    \label{fig:Bfield_theory}}
\end{figure*}

To obtain the oscillator strength with respect to $\sigma^\pm$ polarized light of a given state, we compute its overlap with the zero-momentum state $| \mathbf{k}=0, \pm\rangle = x^\dagger_{\mathbf{k}=0, \pm} |0\rangle$ (see Extended Data Fig.~\ref{fig:Bfield_theory}\textbf{a},\textbf{b}). We then calculate the Zeeman splitting for the optically active umklapp peak. As expected, for weak interactions, the Zeeman splitting of umklapp-scattered excitons is very small~\cite{smolenski2021,shimazaki2021}. However, we find that the magnitude of the Zeeman umklapp splitting increases with stronger exciton-electron interaction potential $|V^\mathrm{X}|$ and decreases for larger electron-hole exchange $J$ (see Extended Data Fig.~\ref{fig:Bfield_theory}\textbf{c}). Most importantly, our calculations reveal that attractive electron-exciton interactions ($V^\mathrm{X} < 0$) lead to a positive Zeeman umklapp splitting, while repulsive interactions ($V^\mathrm{X} > 0$) lead to a negative one. This is consistent with the experimental observation from the main text that the Zeeman splitting for the X, which is a repulsive polaron, and its umklapp are opposite in sign. On the other hand, the AP and WP, which are dominated by attractive electron-exciton interactions, exhibit Zeeman splittings of the same sign. We emphasize that our magnetic-field calculations include electron-exciton interactions only on a mean-field level, which is not sufficient to capture the AP and WP formation. As such, we do not expect to quantitatively describe the Zeeman splitting of the WP and its complex dependency as a function of density $n_\mathrm{e}$. Nevertheless, we qualitatively explain the sign of the Zeeman splitting for the WP. Understanding the details of exciton-polaron formation in a magnetic field is an interesting question for future theoretical work. 

\subsection*{14. Theory: Discussion of singlet and triplet AP}

In Methods Sec.~11, we analyzed polaron formation in the presence of a Wigner crystal, demonstrating how an AP and the associated WP form. So far, our discussion has focused on a single AP and its corresponding WP. However, experimentally, two distinct resonances are observed: the singlet and triplet attractive polaron (AP$_\mathrm{S}$ and AP$_\mathrm{T}$) and their associated Wigner crystal polarons (WP$_\mathrm{S}$ and WP$_\mathrm{T}$). As discussed in the main text, they form due to interactions of the exciton with electrons in the same and opposite valleys, respectively. Here, we discuss the challenges that arise in theoretically modeling the formation of these two distinct AP and WP. 
To first approximation, the formation of singlet and triplet AP can be treated as independent problems. Within this approximation, the model of Methods Sec.~11 applies to both cases separately, once the experimentally determined AP energies are used as input. Despite its simplicity, this approach already captures the density dependence of the WP splitting $\Delta E_{\mathrm{WP}}$ for both singlet and triplet, in good agreement with experiment (see Fig.\ref{fig:Fig_theory}\textbf{g}), thereby justifying this treatment.

A naive extension is to incorporate spin-dependent electron interactions. In this approximation, the energy difference between the singlet and triplet AP is modeled by assuming different interaction strengths between excitons and electrons in the same and opposite valleys. 
However, we find that Chevy-type calculations based on this assumption fail to reproduce the experimental results: both the binding energy and spectral weight of the singlet AP are strongly overestimated. This indicates that introducing spin dependence in the exciton–electron interaction alone is insufficient to explain the observed singlet and triplet AP.

Importantly, recent work in Ref.~\cite{christianen2025asymmetrictrionsmonolayertransition} highlights the significant role the internal structure of the exciton has in accurately capturing the formation of the singlet and triplet trions. They argue that the measured energies of AP$_\mathrm{S}$ and AP$_\mathrm{T}$ can only be explained by taking the non-trivial internal structure of exciton and trion wave functions into account. Such a treatment is beyond our model, where we treat the exciton as a tightly-bound point-like particle, whose wave function has $s$-wave character. Incorporating the internal structure of the exciton into our calculations is a challenging theoretical task and requires a purely electronic description of the exciton formation~\cite{pichler2025purelyelectronicmodelexcitonpolaron}, which is an interesting open problem for future work. 

While our simplified framework cannot fully account for the singlet–triplet AP splitting, it does capture the formation of the WP, which is the central focus of this work. We remark that the appearance of the two AP branches and the formation of the WP share a common microscopic origin but one is not the cause of the other. Both phenomena stem from the band structure of WSe$_2$, which features a more dispersive top conduction band and a heavier, less dispersive lower conduction band. The appearance of the triplet AP branch, in addition to the usual singlet branch, can be traced back to the fact that the bright exciton forms with an electron in the top conduction band, as explained in the main text. These excitons then interact with electrons in the lower, less dispersive, conduction band in the same and opposite valley to form the AP$_\mathrm{S}$ and AP$_\mathrm{T}$, respectively. At the same time, the large effective mass of electrons in the lower conduction band stabilizes the Wigner crystal, which is crucial for WP formation. This contrasts previous studies on MoSe$_2$~\cite{smolenski2021}, where the WC was less stable and no WP was observed. 

%

\newpage

\end{document}